\documentclass[preprint]{elsarticle}
\biboptions{longnamesfirst,square,semicolon}

\usepackage{graphicx}
\usepackage{float}
\usepackage{arydshln}
\usepackage{physics}
\usepackage{xcolor}
\usepackage{longtable}
\usepackage{booktabs}

\usepackage{amssymb}
\usepackage{amsmath}
\usepackage{lineno,hyperref}
\modulolinenumbers[5]

\journal{Journal of Quantitative Spectroscopy \& Radiative Transfer}

\begin{document}

\begin{frontmatter}



  \title{Excited state quantum phase transitions in the bending spectra of molecules}


\author[UHU]{Jamil Khalouf-Rivera}
\ead{yamil.khalouf@dci.uhu.es}
\author[UHU,IC1]{Francisco P\'erez-Bernal}
\ead{francisco.perez@dfaie.uhu.es}
\author[UHU,IC1]{Miguel Carvajal}
\ead{miguel.carvajal@dfa.uhu.es}

\address[UHU]{Dpto. Ciencias Integradas, Facultad de Ciencias
  Experimentales, Centro de Estudios Avanzados en F\'{\i}sica,
  Matem\'aticas y Computaci\'on, Unidad Asociada GIFMAN, CSIC-UHU,
  Universidad de Huelva, Spain} \address[IC1]{Instituto Universitario
  Carlos I de F\'{\i}sica Te\'orica y Computacional, Universidad de
  Granada, Spain}

\begin{abstract}
  We present an extension of the Hamiltonian of the two dimensional limit of the
  vibron model to encompass all possible interactions up to four-body
  operators. We apply this Hamiltonian to the modeling of the bending spectrum
  of four molecules: HNC, H\(_2\)S, Si\(_2\)C, and NCNCS. The selected molecular
  species include linear, bent, and nonrigid equilibrium structures, proving the
  versatility of the algebraic approach which allows for the consideration of
  utterly different physical cases within a single Hamiltonian and a general
  formalism. For each case we compute predicted bending energies and wave
  functions, that we use to depict the associated quantum monodromy diagram,
  Birge-Sponer plot, and participation ratio. In nonrigid cases, we also show the
  bending energy functional obtained using the coherent --or intrinsic-- state
  formalism.
\end{abstract}

\begin{keyword}
  Nonrigid molecules \sep linear molecules \sep excited state quantum phase transition \sep bent molecules  \sep bending rovibrational structure  \sep algebraic Vibron Model 
  


\end{keyword}

\end{frontmatter}


\section{Introduction} 
\label{Intro}

The two-dimensional nature of the vibrational bending degree of freedom,
des\-pi\-te having the linear and bent molecular equilibrium structures physical
limits, implies also rovibrational couplings in quasilinear systems that, even
for triatomic systems, have been the source of frequent misunderstandings in the
description of molecular bending dynamics \cite{Quapp1993}. If the potential
energy surface associated with a particular system has its minimum in the origin
(i.e. it coincides with the molecular axis) the system is said to be linear. If
this minimum is replaced by a maximum, and the potential minimum is located
somewhere else, the molecule is said to have a bent equilibrium structure. Of
course, this is not always so simple --even for textbook examples with a linear
configuration \cite{Jensen2020}-- and, apart from the two well-defined limiting
cases, one often has to deal with quasilinear molecules, whose bending dynamics
is characterized by large amplitude nuclear displacements and are not well
described within the traditional normal mode approach. The possible
cases occurring for intermediate situations can be clearly illustrated by correlation energy diagrams that follow the
evolution of energy levels from one limiting case to the other
\cite{Thorson1960, herz3}.

For quasilinear molecular species, we introduce in the present work a further
distinction between \emph{quasilinear} and \emph{nonrigid} molecules. The
quasilinear case has a molecular bending potential with a flat minimum at the
origin, and its bending spectrum has peculiar signatures, e.g., a positive
anharmonicity in the Birge-Sponer plot or an anomalous ordering of the energy
levels --with maximum vibrational angular momentum levels at lower energies for
a given number of quanta of vibration. The nonrigid case is even richer in
spectroscopic signatures, and it happens in a system with a potential minimum
that is not in the origin, once the bending excitation energy reaches values
high enough to allow for the exploration by eigenstates of the linear
configuration, which in principle is classically forbidden due to the existence
of the barrier to linearity. This explains the switch between negative and
positive anharmonicities in the Birge-Sponer plot that characterizes these
molecules, the well-known Dixon dip \cite{Dixon1964}. Therefore, whenever
vibrational bending levels straddle the barrier to linearity, wavefunctions have
significant components in both the linear and bent regions of configuration
space, giving rise to the particular spectroscopic signatures that characterize
nonrigid spectra. 

The study of large amplitude bending dynamics, and the ensuing coupling between
vibrational and rotational degrees of freedom, has been successfully carried out
making use of different approaches. Most of them solve a zeroth-order
Hamiltonian, where the large amplitude motion (LAM) is placed on equal footing
with rotations, and then consider the complete vibrational-rotational
Hamiltonian with respect to a configuration of reference.  Perfect examples of
this philosophy are the bender Hamiltonian of Hougen-Bunker-Johns \cite{HBJ},
its extensions, like the semirigid bender Hamiltonian \cite{Bunker1977} and the
general semirigid bender Hamiltonian \cite{Ross1988}, or the MORBID
\cite{MORBID} model. The consideration of both rotational and vibrational
degrees of freedom makes these models extremely useful tools for the analysis of
molecular spectra, as they allow for the modeling of experimental term values and
the assignment of quantum labels.
 
The barrier to linearity in nonrigid species is often modeled with Mexican-hat
type potentials. Classical mechanics show that such potentials prevent the
definition of a set of globally valid action-angle variables
\cite{Bates1991}. When this situation is brought into the quantum mechanics
realm, the lack of global action-angle variables translates into the
impossibility of finding a unique set of vibrational quantum numbers globally
valid for the system \cite{Child1998, Winnewisser2006}. This phenomenon, called
\emph{quantum monodromy}, is explained by the changes experienced by the system
spectrum as the linear configuration, initially forbidden,  can be explored for
increasing excitation energies. Introduced by Child, quantum monodromy is
characterized by a piling of states around a critical energy value and a
particular dependence of the bending energy levels on the vibrational angular
momentum, evinced in the quantum monodromy diagram \cite{Child1998}. This
feature was soon used as an effective tool for the labeling of highly-excited
energy levels of water in particularly difficult energy regions
\cite{Child1999}. Quantum monodromy signatures have been later found in
other molecular species \cite{Winnewisser2006, Winnewisser2005, Zobov2005, 
  Winnewisser2010, Winnewisser2014, Reilly2015}.

The present work is based on an algebraic approach that treats quantum many-body
systems with \(N\) degrees of freedom in terms of bosonic realizations of the
\(U(N+1)\) Lie algebra \cite{bookalg}; an approach that has been successfully
applied to the modeling of the structure of widely different physical systems:
nuclei \cite{booknuc,Iachello1993}, hadrons \cite{Bijker1994}, and molecules
\cite{book1}. In the latter case, rovibrational excitations in molecules are
treated as collective bosonic excitations called \emph{vibrons}, and the model
is known as the \emph{vibron model}. This approach was originally introduced by
Iachello for the study of the full rovibrational spectrum of diatomic molecular
species \cite{Iachello:81}. Based on the \(U(4)\) Lie algebra, this model was
later extended to model the spectrum of tri- \cite{Ono:1982} and tetratomic
\cite{Iachello1991} molecular species. The simultaneous treatment of all
rotational and vibrational degrees of freedom comes at a cost, and the required
coupling of \(U(4)\) algebras increase the mathematical and computational
complexity of the model. In the present paper, we use the two-dimensional limit
of the vibron model (2DVM), with a \(U(3)\) dynamical algebra, originally
introduced for the study of single and coupled benders \cite{Iachello1996}. This
model provides an effective Hamiltonian able to deal with large amplitude
bending modes and including from the onset couplings with the rotational
projection around the molecule-fixed z-axis. It allows for a simple,
though complete, description of the linear and bent limiting cases, as well as
of the quasilinear and nonrigid regimes \cite{Iachello2003,PBernal2005}. A
thorough description of the model can be found in Ref.\ \cite{PBernal2008}.

A point of particular interest for algebraic models is the study of
ground state quantum phase transitions (QPTs), also called shape phase
transitions, that are zero-temperature transitions between phases associated
with specific configurations of the system ground state. The different phases
are often associated with well-known limits, called \emph{dynamical
  symmetries} \cite{Cejnar2007}. These transitions are non-thermal and are driven through the
variation of one or several Hamiltonian parameters (control parameters). The
study of such transitions can be traced back to the seminal studies of Gilmore
\cite{Gilmore1979} and it has received a great deal of attention in algebraic
models of nuclear structure \cite{Cejnar2009,Casten2009,Cejnar2010}. The
description of the ground state QPT for the 2DVM model is found in
Ref.\ \cite{PBernal2008}, while different aspects of interest about this transition can be
found in Refs.\ \cite{Zhang2010, PFernandez2011, Calixto2012, Calixto2012b,
  Santos2013, Castanos2015}.

More recently, the ground state QPT concept has been extended to encompass
excited states, with the introduction of excited state quantum phase transitions
(ESQPTs). ESQPTs are characterized by a singularity in the energy spectrum due
to the clustering of excited levels at a certain critical energy
\cite{Cejnar2006, Caprio2008}.  This critical point can be reached in two
different ways: by the variation of a Hamiltonian control parameter, within a certain control parameter
range, for a
constant excitation energy  or by the increase of the excitation energy for a
Hamiltonian with constant parameters. ESQPTs have been studied in different quantum systems, e.g., the nuclear
interacting boson~\cite{Cejnar2009}, Jaynes-Cummings~\cite{Fernandez2011b},
kicked-top \cite{Bastidas2014}, Rabi \cite{Puebla2016}, Lipkin-Meshkov-Glick
(LMG)~\cite{Cejnar2009,Fernandez2009,Yuan2012, Kopylov2015, Wang2019a, Wang2019b}, and
Dicke~\cite{PFernandez2011,Fernandez2011b,Brandes2013, Kloc2018} models.

In the molecular case, it was shown that quantum monodromy and its associated
excited levels clustering can be understood as a manifestation of an ESQPT
\cite{PBernal2008} and it can be described with a formulation common to other
many-body systems \cite{Cejnar2006, Kloc2017, Wang2017}. In fact, as the 2DVM is
the simplest two-level bosonic model with a non-trivial angular momentum, it has
been often used to illustrate the occurrence of ESQPTs in algebraic models
\cite{Caprio2008, PBernal2010, PB_Santos2016}. Due to the advances in
experimental techniques that have made feasible to record highly-excited bending
overtones in nonrigid systems, the molecular bending degree of freedom has been
the first quantum system where experimental signatures of ESQPTs have been found
~\cite{Winnewisser2005, Zobov2005} and explained from an algebraic perspective
\cite{Larese2011, Larese2013}. Other systems where experimental access to ESQPTs
has been achieved are superconducting microwave billiards~\cite{Dietz2013} and
spinor condensates~\cite{Zhao2014}. The authors and Santos have also recently
shown clear evidences of a link between the ESQPT formalism and the study of the
transition state in isomerization chemical reactions \cite{KRivera2019}.

The present work can be considered as an extension and an update of the results
presented in \cite{Larese2011} and \cite{Larese2013}, with the main aim of
calculating spectra within the 2DVM with uncertainties close to spectroscopic
accuracy. In these two works, particular bending modes of several molecular
species with different characteristics --linear, quasilinear, nonrigid, and
bent-- were modeled making use of the 2DVM. The selected species are mostly
four- or five-atomic in \cite{Larese2011} and triatomic in \cite{Larese2013}
and their bending rovibrational structure was explained in terms of the most
general 2DVM Hamiltonian up to two-body interactions (besides the water molecule
case, where extra interactions were taken into consideration). We extend the
number of interactions and make use of the most general 2DVM Hamiltonian
including  up to four-body interactions. To illustrate the improved
results achieved with this  extension, we show
results for four molecular species: hydrogen isocyanide (HNC, linear), hydrogen
sulfide (H$_2$S, bent), cyanogen isothiocyanate (NCNCS, nonrigid), and disilicon
carbide (Si$_2$C, nonrigid). We provide in the Supplementary Material section
the predicted values for not yet measured levels.
        
In addition to the calculated spectra and spectroscopic parameters for
the selected species,  we have also
computed the participation ratio \cite{Evers2008} (PR) of the resulting eigenstates
expressed in the two 2DVM bases, associated with the linear and bent
limiting cases \cite{PBernal2008}. The participation ratio is a quantity,
closely linked to the Shannon entropy \cite{Zelevinsky1996}, that has recently 
been shown to detect  a strong localization in
the eigenstates with energies in the vicinity of the ESQPT critical energy in systems that undergo an ESQPT
\cite{PB_Santos2016,Santos2015,Santos2016}. This fact has important effects in
the system dynamics, which have been proved in the vibron model
\cite{Santos2015}, as well as in its two- \cite{PB_Santos2016}, and
one-dimensional limits \cite{Santos2016}.

 Finally, in nonrigid cases, we make use of the coherent or intrinsic state
 formalism \cite{PBernal2008,Dieperink1980,Leviatan1988} to compute an approximation to the system
 bending energy functional.

\section{The two dimensional limit of the vibron model}
\label{sec-model}

The 2D limit of the vibron model, abbreviated as 2DVM, was introduced in
Ref.~\cite{Iachello1996}. Since then, it has been applied to model
molecular bending degrees of freedom due to its general character,
able to encompass the limiting linear and bent molecular
structures besides the interesting situations in-between them. In the
field of molecular structure, the model was applied to different
problems involving bending vibrations: calculation of infrared or Raman line intensities
\cite{PBernal2005, Oss1998, Temsamani1999, SCastellanos2012, Lemus2014,Marisol2020}, definition of an
algebraic force field for bending vibrations \cite{Sako2000},
computation of Franck-Condon factors \cite{Ishikawa2002}, or
characterizing signatures of non-rigidity in energy spectra
\cite{Iachello2003, Larese2011,Larese2013}. Alternative algebraic approaches to molecular structure that try to
get a firmer grasp on the connection to the traditional phase space
approach are also based on the 2DVM \cite{RArcos2018, EFregoso2018,
  EFregoso2018b}.  More recently, inspired by the use of spectroscopic
information to characterize the transition state in isomerization
reactions presented in Ref.~\cite{Baraban1338}, the authors have found
that the 2DVM model is capable of characterizing the transition state
in such reactions, applying this finding to the bond-breaking system HCN-HNC \cite{KRivera2019}.

Specially important for the present discussion are the detailed description of the model provided in
Ref.~\cite{PBernal2008} and  Refs.\
\cite{Larese2011,Larese2013}, where a careful study of many different
benders is presented and the model results are explained under the
prism of the occurrence of an ESQPT in nonrigid cases. Nonrigid molecules can display spectroscopic signatures typical of a bent or
linear configurations, depending on the energy window considered, as their
levels can be either below or above the barrier to linearity. Therefore, they
showcase the expected ESQPT precursors for a finite system, once the system overcomes the potential barrier and explores the previously
forbidden linear configuration region of the phase space.

It is worth to mention the extension of the model to situations where two
benders are coupled, which implies a significantly larger computational
complexity \cite{Stransky2014,Stransky2015} and where the obtained results can
be explained from the perspective of QPTs involving two bosonic
fluids \cite{Iachello2008,Iachello2009,PBernal2012,Larese2014,Calixto2014}. Related to
this, it is worth mentioning the use of the 2DVM model in the study of the
spectra of 2D crystals with various lattice geometries \cite{Iachello2015}.

The 2DVM associates a \(U(3)\) dynamical algebra to each bender. The nine
generators of this Lie algebra are built as bilinear products of a creation and
an annihilation operator from the basic bricks for the algebra: two Cartesian
bosons (\(\tau_x,\tau_y\)) and a scalar boson (\(\sigma\)). The system
Hamiltonian is obtained as an expansion in terms of operators with the right
symmetry properties that belong either to the dynamical algebra or to one of its
subalgebras. The interested reader can find a detailed mathematical description
of the model in Refs.~\cite{Iachello1996, PBernal2008}. We provide here some
basic details concerning the bases and Hamiltonians we use in the present work
and we also introduce the participation ratio, a quantity used to analyze the
wave function localization in the different bases. We also outline the intrinsic
state formalism, used to obtain the classical limit of the model in the mean
field approximation.

\subsection{The cylindrical and displaced oscillator bases}

There are two possible subalgebra chains starting from the dynamical
algebra, \(U(3)\), and ending in the system symmetry algebra, \(SO(2)\). The
requirement of having \(SO(2)\) as the symmetry algebra implies angular momentum conservation in the system \cite{frank}.
\begin{equation}\label{U3chains}
\begin{array}{cccccl}
     &         &U(2) &          & &Chain (I)\\
     & \nearrow&     & \searrow & &\\
U(3) &         &     &          & SO(2)&\\
     & \searrow&     & \nearrow & &\\
     &         &SO(3)&          & &Chain(II)\\
\end{array}
\end{equation}
Each one of the possible subalgebra chains is known as a \emph{dynamical
  symmetry} and it provides an analytical solution to the problem: an energy
formula that can be mapped to certain physical cases \cite{
  bookalg,frank}. In addition to this, there is a basis associated with every dynamical
symmetry. We proceed to detail the basis quantum numbers and branching
rules for the two dynamical symmetries at stake.

\paragraph{The cylindrical oscillator basis}
The $U(3)\supset U(2)\supset SO(2)$ chain is known as the cylindrical oscillator
chain and it can be mapped with the linear case. Its states are labeled by quantum numbers \(n\) and
\(\ell\)

\begin{equation}
  \left|\begin{array}{ccccc}
      U(3)&\supset& U(2)&\supset& SO(2)\\
      \left[N\right]   &       & n   &       & \ell
    \end{array}\right\rangle ~~,
\label{cobas}
\end{equation}
\noindent and the associated basis states are denoted as  \(|[N];n^\ell\rangle\). The quantum number
$N$ labels the totally symmetric representation of $U(3)$ and the total number of bound states of the system is a function of $N$. The label $n$
is the vibrational quantum number and $\ell$ is the vibrational
angular momentum. The branching rules in this case are
\begin{align}
  n & =  N, N-1, N-2, \ldots, 0 \nonumber\\
  \ell & = \pm n, \pm (n-2), \ldots, \pm 1 \mbox{ or }0~,~~ (n =
  \mbox{odd or even}) ~~.
\end{align}

This is the most convenient basis to fit  vibrational bending data from linear molecules. We provide in \ref{sec-2-3-1} the matrix elements in this basis of the different operators included in  the 2DVM Hamiltonian.

\paragraph{The displaced oscillator basis}
States in the displaced oscillator chain, associated with bending vibrations in molecules with a bent geometric configuration, are characterized by the quantum numbers
\begin{equation}
  \left|\begin{array}{ccccc}
      U(3)&\supset& SO(3)&\supset& SO(2)\\
      \left[N\right]   &       & \omega   &       & \ell
    \end{array}\right\rangle~~.
\label{anosbas}
\end{equation}
and will be denoted as $|[N];\omega,\ell\rangle$.  The branching
rules in this case are
\begin{align}
  \omega & = N, N-2, N-4, \ldots, 1 \mbox{ or }0 ~,~~ (N = \mbox{odd
    or
    even}),\nonumber\\
  \ell & = \pm \omega, \pm (\omega-1) , \ldots , 0~~.
\end{align} 
It is convenient to introduce a vibrational quantum
number \(\nu_b\), which can be identified with the number of quanta of
excitation in the displaced oscillator  \(\nu_b=\frac{N-\omega}{2}\). The branching rules in this case are
\begin{eqnarray}
  \nu_b & = & 0,1, \ldots, \frac{N-1}{2} \mbox{ or }\frac{N}{2}~,~~ (N =
  \mbox{odd or
    even}),\label{nub}\\
 \ell & = & 0, \pm 1,\pm 2, \ldots, \pm (N-2 \nu_b)~~.\nonumber
\end{eqnarray} 
This is the natural basis to fit bending vibration data from  nonrigid and bent molecules.  We provide in \ref{sec-2-3-2} the matrix elements in this basis of the different operators in the four-body 2DVM Hamiltonian.

For  nonrigid molecules, it is expected that low energy
eigenstates would be better defined within the displaced
        oscillator basis set --$SO(3)$ dynamical symmetry-- whereas
        states with energies above the potential barrier should be
        better characterized in the cylindrical oscillator basis set
        --$U(2)$ dynamical symmetry. Therefore, depending on the
        energy, vibrational bending overtones could be assigned either
        to symmetric top quantum labels, \(\nu_b\) and \(K\), or to the 2D harmonic oscillator
        quantum labels, $n$ and $\ell$, used in the linear case. These
        two sets of quantum numbers are linked by the transformation
        \(\nu_b=\frac{n-|\ell|}{2}\) and  \(K = \ell\) (see, e.g. \cite{Quapp1993,
          Winnewisser2006}). 

\subsection{The 2DVM Hamiltonian}

In this work we make use of three different Hamiltonian operators of increasing
complexity. The simplest one, \(\hat {\cal H}\),  has been chiefly used in
the study of ground and excited state QPTs in the 2DVM. It is a very simplified
Hamiltonian that includes the \(\hat n \) operator, from the cylindrical
oscillator dynamical symmetry, and the Pairing opera\-tor \(\hat P\), from the
displaced oscillator dynamical symmetry.

\begin{equation}
  \hat {\cal H} = \varepsilon\left[(1-\xi)\hat n + \frac{\xi}{N-1}\hat P\right]~~.
  \label{modham1}
\end{equation}

The number operator, \(\hat n \), is the total number of $\tau$ bosons and has a
direct physical interpretation in the cylindrical oscillator dynamical symmetry
as the number of quanta of excitation in the linear limit. In fact, it is
defined as \(\hat n = \hat n_x + \hat n_y\) for a two-dimensional harmonic
oscillator. The Pairing operator interpretation is not so direct, it is defined
as \(\hat P=N(N+1)-\hat W^2\), where the first contribution is constant and it
is used for convenience to make the ground state lie at zero energy. The second
term, \( \hat W^2\), is the Casimir operator of the \(SO(3)\)
subalgebra. Therefore it is a squared angular momentum which, nevertheless,
should not be mistaken with the physical angular momentum, that is the
vibrational angular momentum, \(\hat \ell\).  The spectrum of the Pairing
operator is anharmonic, its associated potential has a minimum outside the
origin -thus the displaced oscillator name for the dynamical symmetry,- and the
number of quanta of excitation, \(\nu_b\), can be obtained from the quantum
number \(\omega\) as shown in Eq.~ \ref{nub}. The interested reader can find
more details in Ref. \cite{PBernal2008}.

The Hamiltonian (\ref{modham1}) has two parameters: a global energy scale
\(\varepsilon\) and a control parameter \(\xi\). For \(\xi = 0.0\), the
system is in the first dynamical symmetry (linear limit) while for \(\xi = 1.0\) the
system is in the second one (bent limit). The different values of the
control parameter \(\xi \in [0,1]\) quantifies the weight of one limit or the
other. This is specially adequate in the characterization of ground and excited
state QPTs. A second-order ground state QPT occurs for the critical value
\(\xi_c = 0.2\) \cite{PBernal2008}. In the present work we use this simplified
model Hamiltonian to illustrate the use of the participation ratio as an 
ESQPT probe. In order to perform fits to observed bending spectra, the
Hamiltonian should include more interactions.

\begin{figure}[h]
  \centering
  \includegraphics[width=1.0\textwidth]{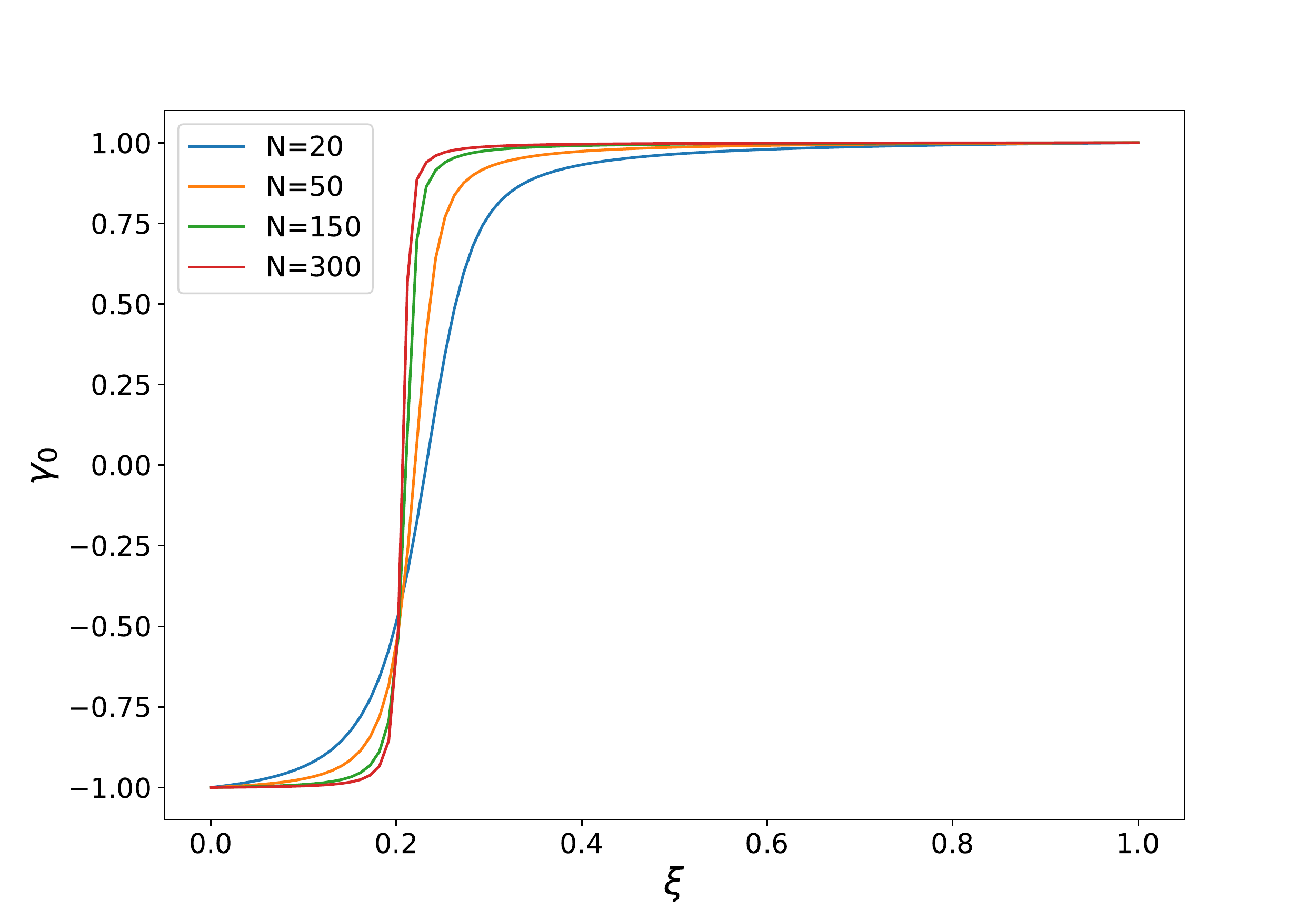}
  \caption{Quasilinearity parameter $\gamma_0$ \cite{Yamada1976} given
    in  Eq.~(\ref{gamma0})  evaluated from the spectrum obtained with model
    Hamiltonian (\ref{modham1}) for control
    parameter values $\xi \in [0,1]$ and different \(N\) values. }
  \label{gamma0_fig}
\end{figure}

It is very illustrative to compute the quasilinearity parameter, \(\gamma_0\), introduced by
Yamada and Winnewisser  \cite{Quapp1993, Yamada1976} computed from the spectrum
of  Hamiltonian (\ref{modham1}) for different values  of the  \(\xi\) control
parameter. The  \(\gamma_0\) parameter is defined as
the ratio of two energy gaps and aims to locate a particular molecule between
the semirigid linear (\(\gamma_0\) = -1) and bent  (\(\gamma_0\) = 1) limits
\begin{equation}
  \label{gamma0}
  \gamma_0 = 1 - 4\frac{E(\nu_b = 0, \ell = 1) - E(\nu_b = 0, \ell = 0) }{E(\nu_b = 1, \ell = 0) - E(\nu_b = 0, \ell = 0) }~,
\end{equation}
\noindent where in the bent and nonrigid cases the \(\ell\) label is replaced by
the $K$ one. In order to make a comparison between the two parameters, we have
depicted in Fig.~\ref{gamma0_fig} a correlation diagram that shows the value of $\gamma_0$ as a function
of $\xi$ for $\xi\in [0,1]$ and different system sizes.

As expected, the \(\gamma_0\) value varies from -1 for \(\xi = 0\) to 1 for
\(\xi = 1\) displaying a sudden change around the critical value of
the control parameter, \(\xi_c = 0.2\), where the system spectroscopic features
change from linear to bent through a nonrigid configuration \cite{Iachello2003,
  PBernal2008}. This change becomes more abrupt the larger the system size,
something that is explained by the fact that the true shape phase transition
happens in the mean field limit, i.e.~for large \(N\) values. In fact, the
quasilinearity parameter (\ref{gamma0}) would be a possible order parameter to
characterize the ground state quantum phase transition in the 2DVM. The
finite-size scaling properties of this transition were studied analitically in
\cite{PFernandez2011}.

A second Hamiltonian of interest is \(\hat H_{2b}\), the most general one- and
two-body Hamiltonian of the 2DVM \cite{Iachello1996, PBernal2008}

\begin{equation}
  \hat H_{2b} = E_0+\epsilon {\hat n} + \alpha {\hat n} ({\hat n}+1) + \beta
  {\hat \ell}^2 + A \hat P~~.
  \label{ham2b}
\end{equation}
\noindent The operators \({\hat n}\) and \({\hat n} ({\hat n}+1)\) are the first
and second order Casimir operators of \(U(2)\) algebra in the cylindrical
oscillator chain. Therefore, the operator \({\hat n} ({\hat n}+1)\) is an
anharmonic correction to the linear limit bending vibration. The pairing
operator \(\hat P\), as mentioned above, is the Casimir operator of \(SO(3)\) in
the displaced oscillator chain, whose spectrum can be mapped to a
two-dimensional anharmonic oscillator. The vibrational angular momentum, \(\hat\ell\),
is common to both dynamical symmetries and it is the physical angular momentum
of the two-dimensional system. In fact, in all the cases considered, the angular
momentum is a constant of the motion, \(\ell\) is a good quantum number, and the
Hamiltonian matrix is block diagonal in \(\ell\). This fact simplifies numerical
calculations, reducing matrix dimensions. This reduction is further
increased because for \(\ell \ne 0\) only positive angular momentum values are
considered. This is explained because, in absence of symmetry-breaking external fields, the first
order angular momentum operator \(\hat\ell\) is not included in the Hamiltonian
and there  positive and negative \(\ell\) value levels are degenerate.

The third Hamiltonian considered is \(\hat H_{4b}\), the most
general 1-, 2-, 3-, and 4-body Hamiltonian expressed as follows
\begin{align}
  \hat H_{4b} =& P_{11} \hat n \nonumber\\
  & + P_{21} \hat n^2 + P_{22} \hat \ell^2 + P_{23} \hat W^2 \nonumber\\
  & + P_{31} \hat n^3 + P_{32} \hat n \hat \ell^2 + P_{33} (\hat n \hat W^2 + \hat W^2 \hat n) \label{H4b}\\
  & + P_{41} \hat n^4 + P_{42} \hat n^2 \hat \ell^2 + P_{43} \hat \ell^4 + P_{44} \hat \ell^2 \hat W^2 \nonumber\\
  & + P_{45} (\hat n^2 \hat W^2 + \hat W^2 \hat n^2) + P_{46} \hat W^4
  + P_{47} (\hat {W}^2 \hat {\overline{W}}^2 + \hat {\overline{W}}^2
  \hat W^2)/2~.\nonumber
\end{align}

This Hamiltonian has fourteen spectroscopic constants \(P_{ij}\), where the
subindexes indicate that the parameter is the \(j\)-th operator among the
\(i\)-body interactions. The operators have been conveniently symmetrized when
they involve products of non-commuting operators. The physical interpretation of
the role of the three- and four-body operators in this Hamiltonian is still
quite clear: \(\hat n^3 \) and \(\hat n^4\) are further anharmonic resonances in
the linear limit; \( \hat \ell^4\) is a centrifugal correction; and \( \hat
W^4\) is an anharmonic correction to the displaced oscillator (bent limit). The
operators \(\hat n \hat \ell^2\) and \(\hat n^2 \hat \ell^2\) are vibration
rotational terms in the linear limit, as well as \(\ell^2 \hat W^2\) for the
bent limit. Finally, \(\hat n \hat W^2 + \hat W^2 \hat n\) and \( \hat n^2 \hat
W^2 + \hat W^2 \hat n^2\) are resonances mixing the two limits of the model and
the term \(\hat {W}^2 \hat {\overline{W}}^2 + \hat {\overline{W}}^2 \hat W^2\)
is a resonance that includes the two possible \(SO(3)\) subalgebras in the
model. This last parameter has only been included for completeness as it has not
been found necessary in any of the fits for the different molecules considered
in this work. The same happens for the \(\hat n^3\) and \(\hat n^4\) operators.

From the matrix elements of the creation and annihilation \(\sigma\) and
\(\tau\) bosons in the two bases associated with the model dynamical symmetries,
published in \cite{PBernal2008}, the operator matrix elements of all operators
in Eq.~(\ref{H4b}) can be derived. We provide the matrix elements of the operators in
the two bases of interest as an appendix to the present work.

\subsection{ESQPT and Participation Ratio}

It has been recently shown that in all vibron model limits the
ESQPT occurring between the $U(N-1)$ and $SO(N)$ dynamical symmetries
(for $N=2,3,4$) implies a strong localization of the wave function
for the state(s) closer to the critical energy of the transition when
expressed in the $U(N-1)$ basis. This fact has important consequences
in the system dynamics \cite{PB_Santos2016,Santos2015,Santos2016}. A
convenient quantity to reveal this localization is the
PR \cite{Evers2008}. This quantity is also named
inverse participation ratio \cite{Izrailev1990} or number of principal
components \cite{Zelevinsky1996}.  If we consider a basis
$\left\{\ket{\psi_i }\right\}_{i=1}^{dim}$, we can express the
eigenstates of our problem as
$\ket{\Psi_k} = \sum^{dim}_{i=1} c_{ki} \ket{\psi_i}$, with the usual
normalization $\sum^{dim}_{i=1} c_{ki}c_{k'i}^* =\delta_{k'k} $. The
PR is defined as the inverse of the sum of $c_{ki}c_{ki}^*$ squared:
\begin{equation}
  PR[\Psi_k]=\frac{1}{\sum_{i=1}^{dim} \left|c_{ki}\right|^4}~.
\end{equation}
\noindent The minimum value of the PR for a given state is one, when
the eigenfunction corresponds exactly with a basis state and all
\(c_{ki}\) components are zero besides one which is equal to
unity. This entails a maximum localization in the selected
basis. On the other hand, the maximum value of the PR is $dim$, the
dimension of the Hamiltonian block. This value is attained once the
wavefunction has all their components equal and non-zero. In the
present case we can express the eigenstates in one of the two bases
associated with the \(U(3)\) dynamical symmetries
\begin{equation}
  \begin{array}{cccl}
    &  & \ket{\Psi_k^{(\ell)}}=\sum_{n} c_{kn}^{(\ell)} \ket{[N]; n^\ell} ~~ \text{ (Cylindrical oscillator chain)}\\
    & \nearrow & \\
    \ket{\Psi_k^{(\ell)}} &  & \\
    & \searrow &  & \\
    & & \ket{\Psi_k^{(\ell)}}=\sum_{\nu_b} d_{k\nu_b}^{(\ell)} \ket{[N]; \nu_b \ell} ~~ \text{ (Displaced oscillator chain)}.
  \end{array}
  \label{PR}
\end{equation}

\begin{figure}[h]
  \centering
  \includegraphics[width=1.0\textwidth]{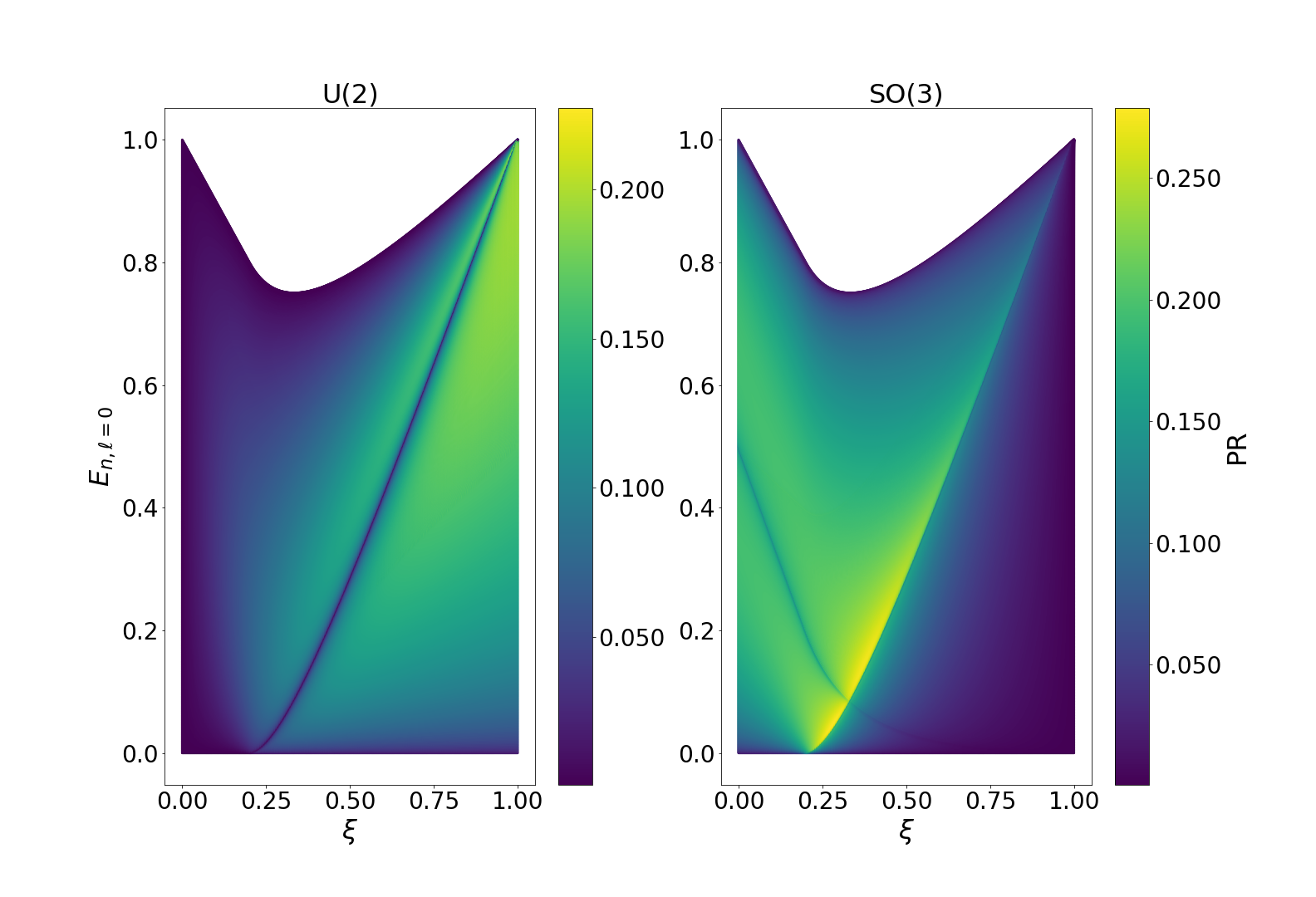}
  \caption{Both panels represent the normalized excitation energies for  $\ell=0$ states with $N=2000$ versus the $\xi$ parameter of the
    model Hamiltonian (\ref{modham1}). Each energy point is colored in
    accordance with the value of the normalized 
    PR for the corresponding eigenstate expressed in the cylindrical oscillator (left panel) or the displaced oscillator (right panel) basis. }
  \label{IPR_1000}
\end{figure}

PR values are usually normalized, dividing the value obtained from
Eq.~(\ref{PR}) by the dimension of the space. This facilitates comparing results for systems with different sizes. The results obtained for
the model Hamiltonian (\ref{modham1}) help to illustrate the
information provided by the PR quantity.  The ground state QPT for the
model Hamiltonian happens for the critical control parameter value
$\xi_c =0.2$ and the ESQPT occurs for control parameters values larger
than \(\xi_c\). The
ESQPT is marked by a nonanaliticity of the energy level density at the
critical energy in the thermodynamic or mean field limit
($N\rightarrow \infty $). In Fig.~\ref{IPR_1000} we depict the
correlation energy diagram for \(N = 2000\) and \(0\le\xi\le 1\) and
we plot the PR for the cylindrical (left panel) and displaced
(right panel) oscillator basis as a heat map. The left panel shows the
high localization of the states on the separatrix that marks the ESQPT
(low PR values) when expressed in the cylindrical oscillator
basis. States located above the separatrix have a $U(2)$ character
--being closer to a linear configuration-- while states below the
separatrix have a $SO(3)$ character -- and are closer to a bent configuration. 
One has to take into account that the PR is a probe too
coarse once the system gets far enough from the dynamical symmetries. As illustrated in
the left panel of Fig.~\ref{IPR_1000}, the minimum PR
value that appears for the
$U(2)$ dynamical symmetry allows for a precise determination of the critical
energy of the ESQPT. However, once the system lies far from a dynamical symmetry,
it is hard to compare the values obtained for the two bases as in both cases
there is a substantial mixing.
    
\subsection{Mean field limit of the 2DVM}
\label{subsec-Vr}

The zero temperature ground and excited state QPTs truly occur in the
thermodynamic limit --or mean field limit-- of the system, for large
system sizes (large $N$ values). In any case, the derivation of a
classical energy functional from the algebraic Hamiltonian is of great
help in understanding and classifying these phenomena. A classical
energy functional, within a \(1/N\) approximation, can be obtained
using the coherent or intrinsic state formalism. This formalism,
originally introduced by Gilmore \cite{Gilmore1979}, was applied in the
first instance to algebraic models in nuclear physics
\cite{Dieperink1980}, and it was later extended to molecular systems
\cite{Leviatan1988}. We present here the basic results, further
details about the intrinsic state formalism results for the 2DVM model
can be found in Ref.\ \cite{PBernal2008}.

The initial step is the consideration of the coherent (or intrinsic) ground state
\begin{equation}
  \label{cond_st}
  \ket{[N];\textbf{r}}= \frac{1}{\sqrt{N!}}\left(b_c^{\dagger}\right)^N\ket{0},
\end{equation}
that is normalized, and where $\textbf{r}$ stands for the 2D classical coordinates and $b_c^{\dagger}$  is the boson condensate operator
\begin{equation}
  b_c^{\dagger}=\frac{1}{\sqrt{1+r^2}}\left[\sigma^{\dagger}+x\tau_x^{\dagger}+y\tau_y^{\dagger}\right].
  \label{bcond}
\end{equation}

Calculating the expectation value of the Hamiltonian (\ref{H4b})
in  the coherent state (\ref{cond_st}) we obtain the system energy functional $E(r)$
\begin{equation}
  E(r)=\bra{[N];\textbf{r}}\hat{H}_{4b}\ket{[N];\textbf{r}}
\end{equation}

The results for the different terms composing the four-body algebraic
Hamiltonian (\ref{H4b}) can be found in \ref{app_coherent}.

The one- and two-body Hamiltonian phase diagram was studied in
\cite{PBernal2008} and it implies a single control parameter and a second order
ground state phase transition between the linear and bent limits, as expected
\cite{Cejnar2007}. The role of the anharmonicity was studied in
\cite{PBernal2010}. The inclusion of three- and four-body operators in
Hamiltonian implies a significantly more complex phase diagram and we are
currently working in its characterization. Once this task is accomplished we
will have a number of essential control parameters, \(\xi_1, \xi_2, \ldots\) and
the correlation between them and the quasilinear parameter (\ref{gamma0}) can be
worked out.

The complete classical limit of the system is obtained considering a complex
variational parameter in the boson condensate (\ref{bcond}). The real and
imaginary parts of the variable are mapped to coordinate and momenta of the
system. We perform a simpler transformation, with a real \(r\) parameter to
obtain the system energy functional. The comparison of this energy functional to
the bending potentials used in configuration space is far from direct. As
mentioned above, the coherent approximation is valid only up to a $1/N$-order.
Considering the $N$ values involved in the study of molecular benders this is a
significant uncertainty. It is possible to go beyond the mean field limit
\cite{PFernandez2011}, but before embarking in this procedure one should grapple
with two other issues that also hinder the above mentioned comparison. The first
one is that the kinetic energy obtained in the intrinsic state approach is
position-dependent \cite{Leviatan1988} and, therefore, it is not equivalent to
the usual kinetic energy operator. Furthermore, one has to deal with the
transformation from the dimensionless variable $r$ to a physical coordinate
measuring the deviation of linearity angle, $\theta$, which implies a connection
between the physical system and its algebraic realization. A linear
approximation to this problem has been previously worked out for the two
dynamical symmetries and extended to situations in-between \cite{Iachello2003,
  Larese2011}. In the present case, the use of higher-order terms in the
Hamiltonian further complicates this connection. In spite of these drawbacks,
and notwithstanding that the results obtained should in principle be considered of a
qualitative nature, the energy functionals resulting from the coherent state
approach provide a fairly intuitive grasp into the model results that helps to
overcome its abstract character. Therefore, in the nonrigid molecules studied,
we provide the resulting energy functionals to help in the interpretation of the
obtained results.

\section{Results}

The main advantage of the 2DVM is the
possibility to encompass, in a computationally simple approach, the full gamut
of behaviors expected for molecular bending vibrations: linear or bent
semi-rigid configurations and the nonrigid case. The latter one is characterized
by a large amplitude, highly anharmonic, degree of freedom and its modeling is
only achieved using a  Hamiltonian
operator that combines interactions from the linear and bent dynamical symmetries. 

In order to bring to light the 2DVM versatility, we model bending vibrational data
for four molecules which have very different spectroscopic signatures: one
semirigid linear (HNC), one semirigid bent (H$_2$S), and two nonrigid molecules
(Si$_2$C and NCNCS) with a large amplitude bent-to-linear mode. In all cases,
the bending spectrum has been reproduced making use of the most general 4-body
Hamiltonian (\ref{H4b}).  For this purpose, we have collected the available data
for the bending degree of freedom under study. The 2DVM model only deals with
vibrational bending levels, therefore it is necessary that the experimental
ro-vibrational term values are fitted and assigned making use of a model able to
extract the vibrational origins from the rotational spectra for each vibrational
band. This vibrational origins are the data we used as a reference for our
model, though in the H$_2$S case (see Subsection \ref{subsec_H2S}) we include in our
calculation the effect of the rotational structure in order to optimize the
agreement with the reported energy values.

Making use of this information, the $P_{ij}$ spectroscopic parameters of
Hamiltonian (\ref{H4b}) have been optimized to reproduce the reported bending spectra,
obtaining in all cases a good agreement between our results and reported
data. We have developed a Fortran source code for the calculation of the
algebraic Hamiltonian (\ref{H4b}) eigenvalues and eigenstates, as well as the
different quantities included in the present work. The code makes use of the
LAPACK \cite{laug} and LAPACK95 \cite{LAPACK95} libraries for matrix
diagonalization and it also performs the requested state assignment tasks. The
parameter optimization procedure is a nonlinear least square fitting using the
\emph{Minuit} Fortran code \cite{minuit}. In every molecule we start from the
two-body minimal Hamiltonian (\ref{ham2b}), fixing to zero the three- and
four-body operators. This Hamiltonian has been previously used in the modeling
of three out of the four cases under study \cite{Larese2011,Larese2013}. After that,
the effects of  different combinations of three- and four-body parameters are
studied, including them in the minimization and using the statistical information provided by
Minuit to choose a minimal set of physically relevant parameter. The minimization
finishes once convergence is reached for a set of parameters and a careful
check of the obtained results has been carried out, to ascertain their physical
character. As can be seen in Tab.~\ref{tab_results}, not all possible $P_{ij}$
parameters in the general algebraic Hamiltonian (\ref{H4b}) have been used. The
code is available under request to the authors and it will be published in a
forthcoming work.

Apart from the $P_{ij}$ spectroscopic parameters, there is an extra integer
parameter, the total number of bosons $N$. This parameter has not been
included in the \emph{Minuit} minimization. Instead, we have manually adjusted
it to optimize the agreement with reported data following one of the methods published in the appendix of \cite{Larese2011}.

The
agreement between calculated ($\{E^{calc}_k\}_{k=1}^{N_{data}}$) and reported results
($\{E^{ref}_k\}_{k=1}^{N_{data}}$) is assessed using the $rms$, defined as follows

\begin{equation}
  rms = \sqrt{\frac{\sum_{k=1}^{N_{data}}\left(E^{calc}_k-E^{ref}_k\right)^2}{N_{data}-n_p}}~,
    \label{rms}
\end{equation}

\noindent where $n_p$ is the number of free parameters in the optimization
routine. 

To facilitate the comparison between the different cases, we report in
Tab.~\ref{tab_results} the optimized parameters in the four cases studied as
well as their one \(\sigma\) confidence interval. As mentioned above, not all
possible $P_{ij}$ parameters in the general algebraic Hamiltonian (\ref{H4b})
have been used.  We have organized this section into four subsections, where we
discuss the results obtained for each molecule. We also include in this table
the value of the total number of bosons, $N$, the $rms$ of the fit, and the
number of vibrational levels involved in the minimization process.

As mentioned above, the present work can be considered as an extension
of previous works where most fits were performed with a two-body
Hamiltonian~\cite{Larese2011,Larese2013}. We have reviewed the results
previously obtained  making use of the four-body
2DVM Hamiltonian (\ref{H4b}) and the agreement in some cases 
have notably improved with the addition of operators of three- or
four-body character, e.g. the HNC case with an  $rms$ that has decreased
from $2.3$ to $0.08$ cm$^{-1}$ with the addition of a single three-body operator.
\begin{table}[h]
  \centering
  \caption{Optimized Hamiltonian parameters ($P_{ij}$, in cm$^{-1}$
    units) for the selected bending degree of freedom of HNC, H$_2$S,
    Si$_2$C, and NCNCS. Values are given with their associated
    uncertainty (one \(\sigma\) confidence interval) between
    parentheses in units of the last quoted digits. The total vibron
    number, $N$, the $rms$ obtained in the fit (in cm$^{-1}$
    units), and the number of data entering the minimization algorithm,  \(N_{data}\), are also included.}
  \begin{tabular}{|c|c|c|c|c|}
    \hline
    Molecule &   HNC           & H$_2$S$^a$ & Si$_2$C     & NCNCS        \\ 
                                                               
    $P_{11}$  & $1414.0(4)$       &-          &  $63.8(5)$    &$331.97(8)$     \\
    $P_{21}$  & $-29.837(15)$     &-          &  $-0.108(18)$ &$-2.0954(6)$      \\
    $P_{22}$  & $15.81(10)$       &$2.897(13)$  &  $0.98(5)$    & $1.190(8)$     \\
    $P_{23}$  & $-8.054(3)$       &$-3.0113(12)$&  $-0.8117(17)$&$-0.58578(17)$    \\ 
    $P_{32}$  & $4.9(10)\times 10^{-2}$ & -         &     -       & -               \\
    $P_{42}$  & -               & -         &  -          & $-2.65(20)\times 10^{-5}$   \\
    $P_{43}$& -                 & $-5.7(3)\times 10^{-05}$ &    -        & -               \\
    $P_{44}$& -                 &$1.235(9)\times 10^{-04}$ &    -        &-                \\
    $P_{46}$& -                 & $1.924(4)\times 10^{-05}$ &   -         & $3.48(8)\times 10^{-7}$      \\
    \hline                                                         
    $N$        & $40$              &$121$       & $49$          &  $150$           \\
    \hline                                                         
    $rms$ &  $0.08$              & $0.20$         &$1.48$        &  $0.79$             \\        
    \(N_{data}\)& $19$            &   $96$        & $37$          & $88$               \\
    \hline
	\end{tabular}
   \flushleft
            {\footnotesize
              $^a$ Two additional parameters besides those listed in this table
              are used in this case to account for rotational effects. See
              Subsection \ref{subsec_H2S}.
            }
                \label{tab_results}
	\end{table}

Apart from the calculation of the fit to the spectrum, we have tried to cast
some light upon the dynamical structure of the different molecular systems. We
show in the four cases under study their quantum monodromy diagram, Birge-Sponer
plot, and participation ratio plot.  The quantum monodromy diagrams and the
Birge-Sponer plots include experimental and calculated bending energy values, as
well as the algebraic four-body Hamiltonian model predictions for yet unknown
energy levels. The PR plots include the results for the optimized zero
vibrational angular momentum eigenstates expressed in the cylindrical
(\ref{cobas}) and displaced oscillator (\ref{anosbas}) bases versus the state
energy.  In the two nonrigid cases, we include as insets in the PR panel the
classical energy functional obtained with the coherent state approach to offer a
more intuitive view of the 2DVM results. Tables with the values of predicted
levels can be found in the Supplementary Material.

\subsection{Hydrogen isocyanide, HNC}
\label{subsec_HNC}
        
Hydrogen isocyanide is an isotopomer of hydrogen cyanide and a linear
molecule. From the many experimental ro-vibrational term values in the
literature, we have selected the 19 available pure $\nu_2$ experimental energy
levels~\cite{MellauHNC}. In the case of linear and quasilinear molecules, the
interactions included in the cylindrical oscillator subalgebra chain
(\ref{U3chains}) are the most relevant, although once the molecule starts
departing from a rigidly-linear behavior, interactions like $\hat W^2$, attached
to the displaced oscillator chain in (\ref{U3chains}), are also required. This
is specially true for quasilinear molecules, due to the flatness of the
potential that characterizes systems close to the ground state QPT
\cite{Larese2011, Larese2013}.

We have obtained a sizable improvement in the fit to this molecule with respect
to the results published in Ref.~\cite{Larese2013}; managing to get a decrease
in the {\em rms} from 2.3~cm$^{-1}$ to 0.08~cm$^{-1}$ with the addition of only
one extra interaction: the 3-body term ${\hat n} {\hat \ell}^2$. The
experimental and calculated vibrational energies are reported in
Tab.~\ref{HNC_tb}. All the states are well below the isomerization transition
state energy, which lies around 12,000~cm$^{-1}$~\cite{KRivera2019}.

\begin{figure}[h]
  \centering
  \includegraphics[trim = 180 0 210 0, clip, width=\textwidth]{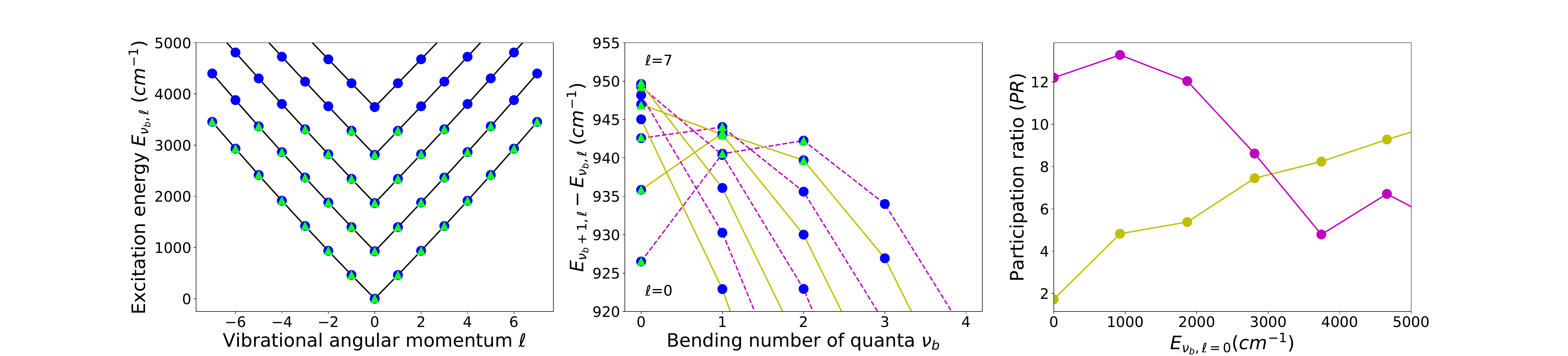}
  \caption{$\nu_2$ bending mode of HNC. Left panel: Quantum monodromy
    plot. Central panel: Birge-Sponer plot. In the left and central
    panels blue circles (green triangles) are calculated
    (experimental) data. Right panel: Participation Ratio of
    $\ell = 0$ eigenstates in the $U(2)$ (yellow line) and $SO(3)$
    (magenta line) bases as a function of the state energy.}
  \label{HNC}
\end{figure}

The calculated energies and eigenstates for the optimized Hamiltonian have been
used in preparing the figures in the different panels of Fig.~\ref{HNC} where
the quantum monodromy plot (left panel) indicates a linear equilibrium
geometrical configuration for this molecule.
However, the Birge-Sponer diagram (center panel of Fig.~\ref{HNC}) is more
complex than expected for a linear molecule, with noticeably different behavior
for states with different vibrational angular momentum values. Besides, the
participation ratio diagram (right panel of Fig.~\ref{HNC}) shows, for
$\ell = 0$ states, a crossing around 3,000~cm$^{-1}$ after which states are more
localized in the $SO(3)$ basis set than in the $U(2)$ basis set. As already
highlighted in Ref.~\cite{Larese2013}, the Birge-Sponer plot indicates that
there is a competition between anharmonicity and pairing operators in the
subspaces with different vibrational angular momenta. The inclusion of the new
cubic term seems to conveniently tackle with this. In any case, the eigenvectors
are more complex than expected for a semirigid linear molecule, as can be deduced from the
crossing of the two curves of the PR plot. This could be due to the influence of
the isomerization barrier for this system. However, it lies at 12,000~cm$^{-1}$,
too far to explain such a low energy feature in the PR~\cite{KRivera2019}. This
case is a good example of why the PR is too simple to ascribe a 
linear or bent character to the wave function if the system is far enough from the
limiting cases defined by the two dynamical symmetries.

%

         \begin{table}[h]
           \centering
           \caption{Experimental and calculated term values and residuals for the  bending mode of HNC. Units of cm$^{-1}$.}
            \begin{tabular}{cccc}
              \hline
              $(n,\ell)^a$ & Exp.$^b$ & Cal.$^c$ & Exp.-Cal.$^d$ \\
              \hline
       (  2  0)    &     926.50     &      926.5071   & -0.007 \\ 
       (  4  0)    &    1867.05     &     1867.0497   & 0.000 \\ 
       (  6  0)    &    2809.29     &     2809.2992   & -0.009 \\ 
       (  1  1)    &     462.72     &      462.6863   & 0.034 \\ 
       (  3  1)    &    1398.56     &     1398.5296   & 0.030 \\ 
       (  5  1)    &    2341.84     &     2341.7558   & 0.084 \\ 
       (  7  1)    &    3281.50     &     3281.4508   & 0.049 \\ 
       (  2  2)    &     936.05     &      936.1066   & -0.057 \\ 
       (  4  2)    &    1878.72     &     1878.6866   & 0.033 \\ 
       (  6  2)    &    2822.75     &     2822.7088   & 0.041 \\ 
       (  3  3)    &    1419.97     &     1419.9198   & 0.050 \\ 
       (  5  3)    &    2366.83     &     2366.9073   & -0.077 \\ 
       (  7  3)    &    3309.78     &     3309.9472   & -0.167 \\ 
       (  4  4)    &    1913.87     &     1913.8403   & 0.030 \\ 
       (  6  4)    &    2863.11     &     2863.1206   & -0.011 \\ 
       (  5  5)    &    2417.57     &     2417.6251   & -0.055 \\ 
       (  7  5)    &    3367.37     &     3367.2552   & 0.115 \\ 
       (  6  6)    &    2930.90     &     2931.0649   & -0.165 \\ 
       (  7  7)    &    3453.78     &     3453.9760   & -0.196 \\ 

                \hline
            \end{tabular}
            \flushleft
            {\footnotesize
              $^a$  Cylindrical oscillator basis quantum labels assigned to the optimized eigenvectors.\\
            $^b$ Experimental energies from Ref.~\cite{MellauHNC}.\\
            $^c$ Calculated energies.\\
            $^d$ Difference between experimental and calculated energies.
            }
            \label{HNC_tb}
        \end{table}

        \subsection{Hydrogen sulfide, H$_2$S}
\label{subsec_H2S}

The rovibrational spectrum of hydrogen sulfide has been exhaustively
studied~(see, e.g., Ref.~\cite{Carvajal2015} and references therein).
Rovibrational energies for bending
overtones are known in bands up to $\nu_2=5$,
inclusive~\cite{Lechuga1984,H2S_1,BROWN1998,ULENIKOV2005,AZZAM2013,Ulenikov2020}. This
is a bent molecule and only interactions associated with the displaced
oscillator dynamical symmetry have been required in order to obtain a good fit.

The H$_2$S molecule is an asymmetric-top and its experimental $\nu_2$ bending
states are therefore labeled as $|\nu_b; J, K_a , K_c \rangle$, where the
quantum numbers $K_a$ and $K_c$ are the projections in the molecular fixed frame
system of the rotational angular momentum $J$ along the $z$-axis and $y$-axis,
respectively (assuming the $I^r$ convention). In this case, one should start by
selecting those states that are more into the 2DVM scope, taking into
account that in the SO(3) dynamical symmetry, the model can be mapped into a 2D
truncated rovibrator, and the $\ell$ angular momentum can be identified with $K$,
the angular momentum projection on the figure axis of the molecule. Therefore,
in the case of asymmetric top molecules, one should consider whether the molecule
is closer to a prolate or oblate rotator. Hydrogen sulfide is closer to the
oblate limit, and therefore we have used as an input for the model
 the 96 available experimental bending rovibrational levels $|\nu_b; J, K_a=0 ,
 K_c=J \rangle$,  with $J=K_c=0,\ldots, 20$ with
\(A_1\) (even $J$) or \(B_1\) (odd $J$) symmetry
(for a discussion on the symmetry of these states see, for example, \cite{Carvajal2015,ALVAREZBAJO2012}). 

In this case, we have added to the
Hamiltonian (\ref{H4b}) two extra operators that are linear in the absolute value of the
vibrational angular momentum, ${|\ell|}$, associated with the spectroscopic
parameters  $B$ and $B_{\nu_b}$.

        \begin{equation}\label{bentHamilt}
          \hat H_{bent}  = 
           \hat H_{4b} + B |\ell| + B_{\nu_b} |\ell| \hat{W}^2 ~~,
        \end{equation}

These two parameters, in particular $B$, are fundamental to understand the
improvement achieved in this case when comparing our results with  the results in
Ref.~\cite{Larese2013}. The need of these extra interaction terms in Eq.~(\ref{bentHamilt})
can be understood considering the linear $J$ term that stems from the rotational
term $J (J+1)$ in the rovibrational Hamiltonian while the operator associated
with the $B_{\nu_b}$ parameter introduces an extra centrifugal correction \cite{Carvajal2015,ALVAREZBAJO2012}.

Our analysis started with the fit of the Hamiltonian~(\ref{H4b}) relevant parameters, those associated with the displaced oscillator
dynamical symmetry (see Tab.~\ref{tab_results}), obtaining an {\em rms} of
11.02~cm$^{-1}$. Once the parameter $B$ is included, the {\em rms} decreases to
0.93~cm$^{-1}$. The final result has an {\em rms} of 0.20~cm$^{-1}$. The
optimized values of the $B$ and $B_{\nu_b}$ parameters are \(B = 18.98(21)\) and
$B_{\nu_b}=-6.29(15)\times 10^{-4}$, both in cm$^{-1}$ units.  The
calculated bending energy levels, shown in Tab.~\ref{H2S_tb}, have a
satisfactory agreement with the experimental data. This agreement is markedly
better than the agreement obtained in Ref.~\cite{Larese2013}, where only 35
experimental term values were included in the fit. Therefore, the present
four-body Hamiltonian, plus the rotational energy correction, achieves a
significant improvement in the optimization and, in our experience, within these
accuracy levels, the predicted spectra can help in the labeling of
not-yet-assigned experimental energy
levels. 

\begin{figure}[h]
  \centering
  \includegraphics[width=1.0\textwidth]{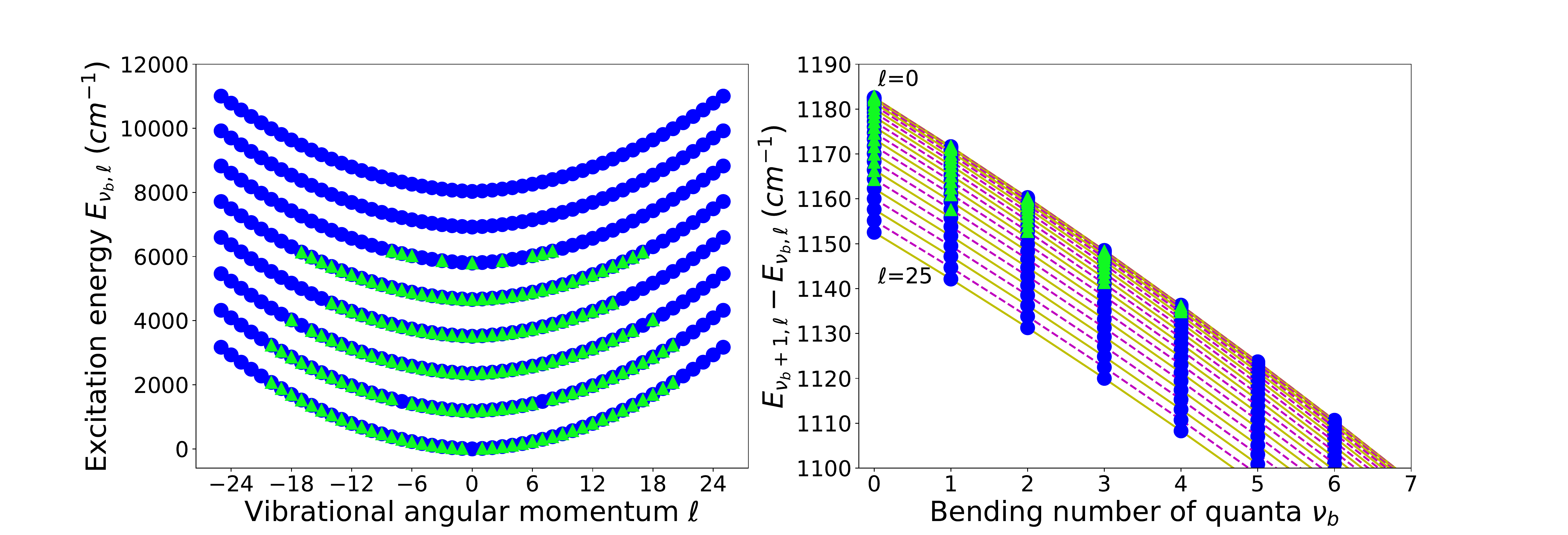}
  \caption{$\nu_2$ bending mode of H$_2$S. Left panel: quantum monodromy
              plot. Right panel: Birge-Sponer plot. In both panels blue
              circles (green triangles) are calculated (experimental) data.}
  \label{H2S}
\end{figure}

The energy term values and eigenvectors obtained from the optimized Hamiltonian
have been used to compute the quantities depicted in the two panels
of Fig.~\ref{H2S}. The quantum monodromy plot (left panel) and Birge-Sponer diagram
(right panel) obtained in this case are textbook examples of a semirigid bent
molecular species. As in this case all the $\hat H_{4b}$ operators included
belong to a dynamical symmetry, we decided not to include the PR. In the $SO(3)$
basis all eigenstates have PR equal to unity and maximal localization, while in
the $U(2)$ case the PR would be given by the transformation bracket between the
two basis.


        \begin{table}[h]
          \centering
          \caption{Experimental and calculated term values and
            residuals for the bending mode of H$_2$S
            with quantum numbers $(\nu_b, J, K_a=0, K_c=J)$ ($A_1$ or $B_1$
            symmetry). Units of cm$^{-1}$.}
            \begin{minipage}{0.4\textwidth}
            \begin{tabular}{cccc |}     

              \hline
              $(\nu_b,K)^a$ & Exp.$^b$ & Cal.$^c$ & Exp.-Cal.$^d$ \\
              \hline
( 0  1) &   13.75 & 14.4158 & -0.66952 \\
( 0  2) &   38.02 & 38.2719 & -0.25581 \\
( 0  3) &   71.42 & 71.5662 & -0.14191 \\
( 0  4) &  114.17 & 114.2953 & -0.12315 \\
( 0  5) &  166.34 & 166.4546 & -0.11116 \\
( 0  6) &  227.94 & 228.0378 & -0.09286 \\
( 0  7) &  298.97 & 299.0375 & -0.06786 \\
( 0  8) &  379.41 & 379.4448 & -0.03755 \\
( 0  9) &  469.25 & 469.2497 & -0.00365 \\
( 0 10) &  568.47 & 568.4406 & 0.03187 \\
( 0 11) &  677.07 & 677.0046 & 0.06701 \\
( 0 12) &  795.03 & 794.9274 & 0.09960 \\
( 0 13) &  922.32 & 922.1934 & 0.12747 \\
( 0 14) & 1058.93 & 1058.7857 & 0.14837 \\
( 0 15) & 1204.85 & 1204.6860 & 0.16002 \\
( 0 16) & 1360.03 & 1359.8745 & 0.15930 \\
( 0 17) & 1524.48 & 1524.3303 & 0.14495 \\
( 0 18) & 1698.14 & 1698.0309 & 0.11375 \\
( 0 19) & 1881.02 & 1880.9526 & 0.06325 \\
( 0 20) & 2073.06 & 2073.0702 & -0.00924 \\
( 1  0) & 1182.58 & 1182.1654 & 0.41456 \\
( 1  1) & 1196.47 & 1196.8248 & -0.35769 \\
( 1  2) & 1220.87 & 1220.8053 & 0.06288 \\
( 1  3) & 1254.26 & 1254.1050 & 0.15850 \\
( 1  4) & 1296.86 & 1296.7204 & 0.14046 \\
( 1  5) & 1348.76 & 1348.6469 & 0.10921 \\
( 1  6) & 1409.96 & 1409.8782 & 0.08236 \\
( 1  8) & 1560.27 & 1560.2244 & 0.04448 \\
( 1  9) & 1649.35 & 1649.3203 & 0.03154 \\
( 1 10) & 1747.71 & 1747.6830 & 0.02310 \\
( 1 11) & 1855.32 & 1855.2998 & 0.01524 \\
( 1 12) & 1972.17 & 1972.1563 & 0.00962 \\
( 1 13) & 2098.24 & 2098.2370 & 0.00245 \\
( 1 14) & 2233.52 & 2233.5248 & -0.00482 \\
( 1 15) & 2377.99 & 2378.0016 & -0.01616 \\
( 1 16) & 2531.62 & 2531.6475 & -0.03001 \\
( 1 17) & 2694.39 & 2694.4416 & -0.05001 \\
( 1 18) & 2866.29 & 2866.3615 & -0.07479 \\
( 1 19) & 3047.28 & 3047.3834 & -0.10789 \\
( 1 20) & 3237.33 & 3237.4822 & -0.14761 \\
( 2  0) & 2353.96 & 2353.4272 & 0.53280 \\
( 2  1) & 2368.02 & 2368.3260 & -0.31093 \\
( 2  2) & 2392.57 & 2392.4289 & 0.14305 \\
( 2  3) & 2425.98 & 2425.7339 & 0.24608 \\
( 2  4) & 2468.45 & 2468.2375 & 0.21325 \\
( 2  5) & 2520.10 & 2519.9351 & 0.16101 \\
( 2  6) & 2580.93 & 2580.8204 & 0.10757 \\
( 2  7) & 2650.94 & 2650.8861 & 0.05828 \\
              \hline
            \end{tabular}
          \end{minipage}\hfill
            \begin{minipage}{0.4\textwidth}
            \begin{tabular}{cccc}
              \hline
              $(\nu_b,K)^a$ & Exp.$^b$ & Cal.$^c$ & Exp.-Cal.$^d$ \\
              \hline
( 2  8) & 2730.14 & 2730.1233 & 0.01338 \\
( 2  9) & 2818.50 & 2818.5219 & -0.02676 \\
( 2 10) & 2916.01 & 2916.0702 & -0.06272 \\
( 2 11) & 3022.66 & 3022.7555 & -0.09498 \\
( 2 12) & 3138.44 & 3138.5634 & -0.12111 \\
( 2 13) & 3263.34 & 3263.4784 & -0.14310 \\
( 2 14) & 3397.32 & 3397.4835 & -0.15999 \\
( 2 15) & 3540.39 & 3540.5604 & -0.17126 \\
( 2 16) & 3692.51 & 3692.6893 & -0.17623 \\
( 2 18) & 4023.85 & 4024.0181 & -0.16597 \\
( 3  0) & 3513.79 & 3513.3476 & 0.44241 \\
( 3  1) & 3528.02 & 3528.4819 & -0.46341 \\
( 3  2) & 3552.76 & 3552.7051 & 0.05449 \\
( 3  3) & 3586.21 & 3586.0153 & 0.19295 \\
( 3  4) & 3628.58 & 3628.4090 & 0.17204 \\
( 3  5) & 3680.00 & 3679.8815 & 0.11853 \\
( 3  6) & 3740.49 & 3740.4267 & 0.06612 \\
( 3  7) & 3810.05 & 3810.0371 & 0.01212 \\
( 3  8) & 3888.67 & 3888.7039 & -0.03842 \\
( 3  9) & 3976.33 & 3976.4169 & -0.08406 \\
( 3 10) & 4073.04 & 4073.1646 & -0.12356 \\
( 3 11) & 4178.78 & 4178.9341 & -0.15506 \\
( 3 12) & 4293.53 & 4293.7111 & -0.17708 \\
( 3 13) & 4417.29 & 4417.4801 & -0.18743 \\
( 3 14) & 4550.04 & 4550.2241 & -0.18406 \\
( 4  0) & 4661.68 & 4661.4963 & 0.18369 \\
( 4  1) & 4676.10 & 4676.8620 & -0.76288 \\
( 4  2) & 4701.06 & 4701.2035 & -0.14727 \\
( 4  3) & 4734.58 & 4734.5187 & 0.06025 \\
( 4  4) & 4776.88 & 4776.8044 & 0.08028 \\
( 4  5) & 4828.12 & 4828.0557 & 0.06006 \\
( 4  6) & 4888.30 & 4888.2666 & 0.03130 \\
( 4  7) & 4957.43 & 4957.4295 & 0.00067 \\
( 4  8) & 5035.51 & 5035.5357 & -0.02751 \\
( 4  9) & 5122.53 & 5122.5750 & -0.04977 \\
( 4 10) & 5218.47 & 5218.5358 & -0.06318 \\
( 4 11) & 5323.34 & 5323.4052 & -0.06383 \\
( 4 12) & 5437.12 & 5437.1691 & -0.04788 \\
( 4 13) & 5559.80 & 5559.8117 & -0.01193 \\
( 4 14) & 5691.37 & 5691.3162 & 0.04929 \\
( 4 15) & 5831.81 & 5831.6642 & 0.14280 \\
( 4 16) & 5981.11 & 5980.8360 & 0.27088 \\
( 4 17) & 6139.26 & 6138.8106 & 0.44855 \\
( 5  0) & 5797.24 & 5797.4504 & -0.21044 \\
( 5  3) & 5870.73 & 5870.8214 & -0.09345 \\
( 5  6) & 6024.00 & 6023.9173 & 0.07835 \\
( 5  7) & 6092.74 & 6092.6406 & 0.10417 \\
( 5  8) & 6170.33 & 6170.1960 & 0.13321 \\
              \hline
            \end{tabular}
            \end{minipage}
            \flushleft {\footnotesize
              $^a$  Displaced oscillator basis quantum labels assigned to the optimized eigenvectors.\\
              $^b$ Experimental energies from \cite{Lechuga1984,H2S_1,BROWN1998,ULENIKOV2005,AZZAM2013,Ulenikov2020}.\\
              $^c$ Calculated energies.  $^d$ Difference between experimental
              and calculated energies.  }
            \label{H2S_tb}
        \end{table}


        \subsection{Disilicon carbide, Si$_2$C}
        \label{subsec_Si2C}
        Disilicon carbide is a floppy triatomic
        molecule~\cite{Kafafi1983,Rittby1991,Presilla1991} which, in recent years,
        has been the subject of a number of experimental works on its
        rotational and rovibrational
        spectra~\cite{Reilly2015,Mccarthy2015} mostly motivated by the
        relevance of silicon and carbon clusters in astronomy and in
        technical applications. The presence of this molecule in IRC + 10216 was confirmed in
        2015 \cite{Cernicharo2015}.
        
        The large amplitude
        motion of Si$_2$C stems from the $\nu_2$ bending mode. The
        available experimental rovibrational term values of the
        excited $\nu_2$ bands, up to $\nu_b=\nu_2=13$ and $\ell=3$
        (approx. up to 1600~cm$^{-1}$) denote a pronounced quantum monodromy
        effect~\cite{Reilly2015}. This dataset has been
        used as an input to optimize the four one- and two-body spectroscopic parameters in Hamiltonian
        (\ref{H4b}). They have been fitted to reproduce a
        total of 37 available experimental data with an {\em rms} of
        1.48~cm$^{-1}$ (see Tab.~\ref{tab_results}).  Note that this
        result is slightly less than the reported experimental uncertainty of
        2~cm$^{-1}$~\cite{Reilly2015} and it can be considered a very good
        agreement with the reported data. It does not seem necessary to include
        higher order interactions in this case and our results can be compared to the results
        reported in previous
        works~\cite{Reilly2015,Koput2017} though taking into account that these calculations are not of a
        phenomenological nature. In both Refs~\cite{Reilly2015,Koput2017},  the
        rovibrational energies are obtained from
        an \textit{ab initio} potential energy surface. The bending energies
        obtained with the present approach,
        labeled with displaced oscillator (bent molecule) quantum
        numbers, are reported in Tab.~\ref{Si2C_tb}, together with
        the reported experimental data values and fit residuals.

\begin{figure}[h]
          \centering %
          \includegraphics[trim = 180 0 210 0, clip,
          width=\textwidth]{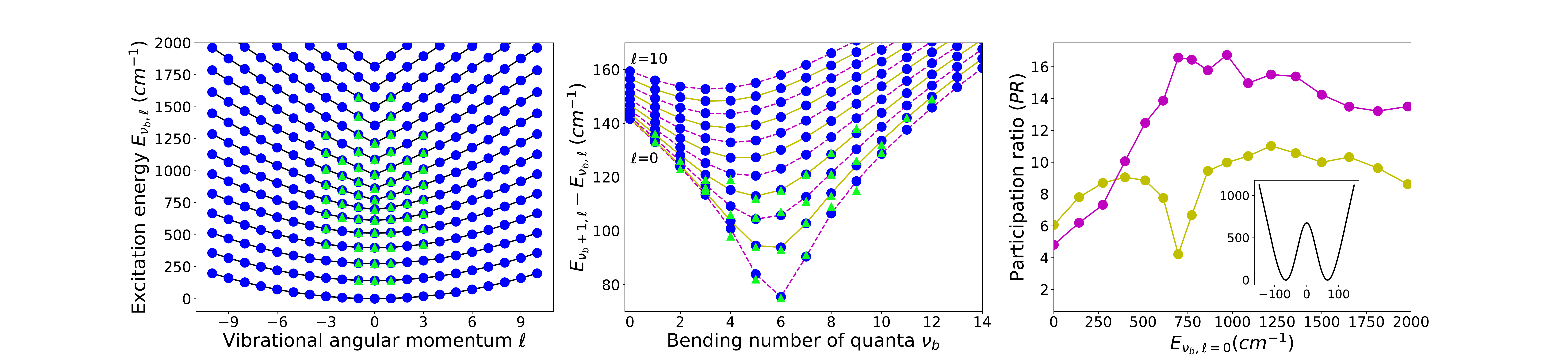}
          \caption{$\nu_2$ bending mode of Si\(_2\)C. Left panel:
            quantum monodromy plot. Central panel: Birge-Sponer plot. In both
            panels blue circles (green triangles) are calculated (experimental)
            data. Right panel: Participation Ratio of $\ell = 0$ eigenstates in
            the $U(2)$ (yellow line) and $SO(3)$ (magenta line) bases as a
            function of the state energy.  The bending energy functional derived
            using the coherent state formalism is included as an inset in the
            right panel.}
          \label{Si2C}
        \end{figure}

        The calculated energies and bending eigenstates have been used
        to compute the different quantities displayed in
        Fig.~\ref{Si2C}. The quantum monodromy diagram --left panel--
        and the Birge-Sponer plot --mid panel-- are in very
        good accordance with the results published in
        Refs.~\cite{Reilly2015,Koput2017}, with a critical ESQPT energy
        and Dixon dip  at $\nu_b=6$, and a barrier to
        linearity that extends up to around 700~cm$^{-1}$. The PR
        results, depicted in the right panel, imply a slightly
        larger eigenstate localization when they are expressed in the
        displaced oscillator basis, and how the trend is reversed for
        higher energy values. In particular, this plot illustrates
        vividly the predicted localization effects in the $U(2)$ basis for the
        $\nu_b=6$ overtone, the closest one to  the
        critical ESQPT energy. As expected, the closest states to the critical
        energy have a very large component in the
        $|n=0^{\ell = 0}\rangle$ basis state \cite{PB_Santos2016,Santos2015,Santos2016}.
        The energy functional is given as a inset in the right panel of Fig.~\ref{Si2C}.
        From the energy functional the
         barrier to linearity value can be estimated to be around $\sim
         675$~cm$^{-1}$. As explained in Subsec.~\ref{subsec-Vr}, the implicit
         $1/N$ errors in the mean field approximation and the difference between the kinetic energy
         operators can explain why the value is too low, when compared with the values 783(48)~cm$^{-1}$ -from the information of the Dixon
        dip~\cite{Reilly2015}- and  832~cm$^{-1}$ obtained using {\em ab initio}
        calculations~\cite{Koput2017}. 

        \begin{table}[h]
          \centering
          \caption{Experimental and calculated term values and
            residuals for the bending mode of Si$_2$C. Units of cm$^{-1}$.
              }
            \begin{tabular}{cccc}
              \hline
              $(\nu_b,K)^a$ & Exp.$^b$ & Cal.$^c$ & Exp.-Cal.$^d$ \\
              \hline
              (1 0) &  140.0 &   141.5931 & -1.593 \\ 
              (1 1) &  142.0 &   143.8421 & -1.842 \\ 
              (2 0) &  273.0 &   274.6069 & -1.607 \\ 
              (2 1) &  278.0 &   277.1851 &  0.815 \\ 
              (3 0) &  399.0 &   398.3725 &  0.627 \\ 
              (3 1) &  401.0 &   401.4841 & -0.484 \\ 
              (3 3) &  425.0 &   425.7597 & -0.760 \\ 
              (4 0) &  515.0 &   511.7915 &  3.209 \\ 
              (4 1) &  516.0 &   515.9391 &  0.061 \\ 
              (4 3) &  544.0 &   546.6778 & -2.678 \\ 
              (5 0) &  613.0 &   612.6117 &  0.388 \\ 
              (5 1) &  622.0 &   619.6988 &  2.301 \\ 
              (5 2) &  636.0 &   637.0290 & -1.029 \\ 
              (5 3) &  663.0 &   661.9059 &  1.094 \\ 
              (6 0) &  695.0 &   696.4984 & -1.498 \\ 
              (6 1) &  716.0 &   714.2134 &  1.787 \\ 
              (6 2) &  741.0 &   741.3982 & -0.398 \\ 
              (6 3) &  775.0 &   774.8380 &  0.162 \\ 
              (7 0) &  770.0 &   771.9460 & -1.946 \\ 
              (7 1) &  809.0 &   808.0725 &  0.928 \\ 
              (7 2) &  848.0 &   847.2296 &  0.770 \\ 
              (7 3) &  890.0 &   890.0296 & -0.030 \\ 
              (8 0) &  861.0 &   862.3533 & -1.353 \\ 
              (8 1) &  912.0 &   910.8638 &  1.136 \\ 
              (8 2) &  959.0 &   959.9006 & -0.901 \\ 
              (8 3) & 1011.0 &  1011.0350 & -0.035 \\ 
              (9 0) &  970.0 &   968.8106 &  1.189 \\ 
              (9 1) & 1025.0 &  1024.8626 &  0.137 \\ 
              (9 2) & 1080.0 &  1081.3839 & -1.384 \\ 
              (9 3) & 1140.0 &  1139.4754 &  0.525 \\ 
              (10 0) & 1085.0 &  1087.2974 & -2.297 \\ 
              (10 1) & 1151.0 &  1149.1694 &  1.831 \\ 
              (10 3) & 1278.0 &  1275.7166 &  2.283 \\ 
              (11 0) & 1214.0 &  1215.9113 & -1.911 \\ 
              (11 1) & 1283.0 &  1282.7567 &  0.243 \\ 
              (12 1) & 1425.0 &  1424.8196 &  0.180 \\ 
              (13 1) & 1574.0 &  1574.7200 & -0.720 \\ 

              \hline
            \end{tabular}
            \flushleft
            {\footnotesize
              $^a$  Displaced oscillator basis quantum labels assigned to the optimized eigenvectors.\\
              $^b$ Experimental energies from~\cite{Reilly2015}.\\
              $^c$ Calculated energies.\\
              $^d$ Difference between experimental and computed energies. 
            }
            \label{Si2C_tb}
        \end{table}

\subsection{Cyanogen isothiocyanate, NCNCS}
\label{subsec_NCNCS}

In this subsection we apply the 2DVM to the $\nu_7$ bending mode of
cyanogen isothiocyanate (NCNCS), a nonrigid molecule~\cite{Ross1988,King1985},
characterized by a large amplitude CNC bending. The spectrum for this mode has
been carefully charted in the microwave and millimeter ranges for several
highly-excited $\nu_7$ states and this molecule has been subject of several
works aiming to study quantum monodromy effects, making it one of the best
examples of quantum monodromy found to date \cite{Winnewisser2006,
  Winnewisser2005, Winnewisser2010, Winnewisser2014}.

Being so rich in
spectroscopic features, the disentangling of the spectra of NCNCS
is a  cumbersome task, and it displays unusual features  in its rotational
and vibrational spectra. In the case of Ref.~\cite{Winnewisser2010} the GSRB
model is used to allow the simultaneous treatment of rotations and vibrations
and the calculation of the vibrational band origins. We use these values as an input to our
model. The simultaneous treatment of rotations and vibrations is possible in the
$U(4)$ based original vibron model \cite{bookmol}, though at the cost of a  more complex
mathematical apparatus than the 2DVM one. However, it is possible to carry out a simpler description of the
rotational spectra for nonrigid molecules within the 2DVM. In this case the
\(\Delta B_{eff}\) parameter, defined as \(\Delta B_{eff} = B(\nu_b, \ell) -
B(0,0)\), that quantifies the rotational constant dependence on the bending
number of quanta and the vibrational angular momentum -or \(K\) value, \(\ell =
K\)- can be expressed as a series expansion in the number operator \cite{Larese2011} 

\begin{equation}
  \Delta B_{eff} = a_1 \hat n  + a_2 \hat n(\hat n +1) + \ldots ~,
  \label{Beff}
\end{equation}
\noindent with \(a_1 >> a_2 >> ...\). 

    In this work we analyze the CNC bending mode band origins  up
    to $\nu_b=\nu_7=5$ and $K_a=20$ reported in~\cite{Winnewisser2010}. These
    data correspond to \(\nu_b, J = K_a, K_a\) levels and were obtained analyzing
    the experimental rovibrational term values with 
    use of the GSRB model (See Tables 8 to 11 in Ref.~\cite{Winnewisser2010}).
     Eighty-eight
    reported band origins  have been fitted using six operators of the four-body
    Hamiltonian~(\ref{H4b}) (full one- and two-body plus two four-body operators), with a final {\em rms}=0.79~cm$^{-1}$
    (see Tab.~\ref{tab_results}), which improves the {\em rms} of
    2.2~cm$^{-1}$ obtained in~\cite{Larese2011} though a direct comparison is
    not easy as in this paper data from \cite{Winnewisser2005} were used in the
    fit. The improvement achieved can be explained from the
    inclusion of the four-body interactions ${\hat \ell}^2 {\hat n}^2$ and ${\hat W}^4$.

\begin{figure}[h]
            \centering
            \includegraphics[trim = 80 0 100 0, clip, width=\textwidth]{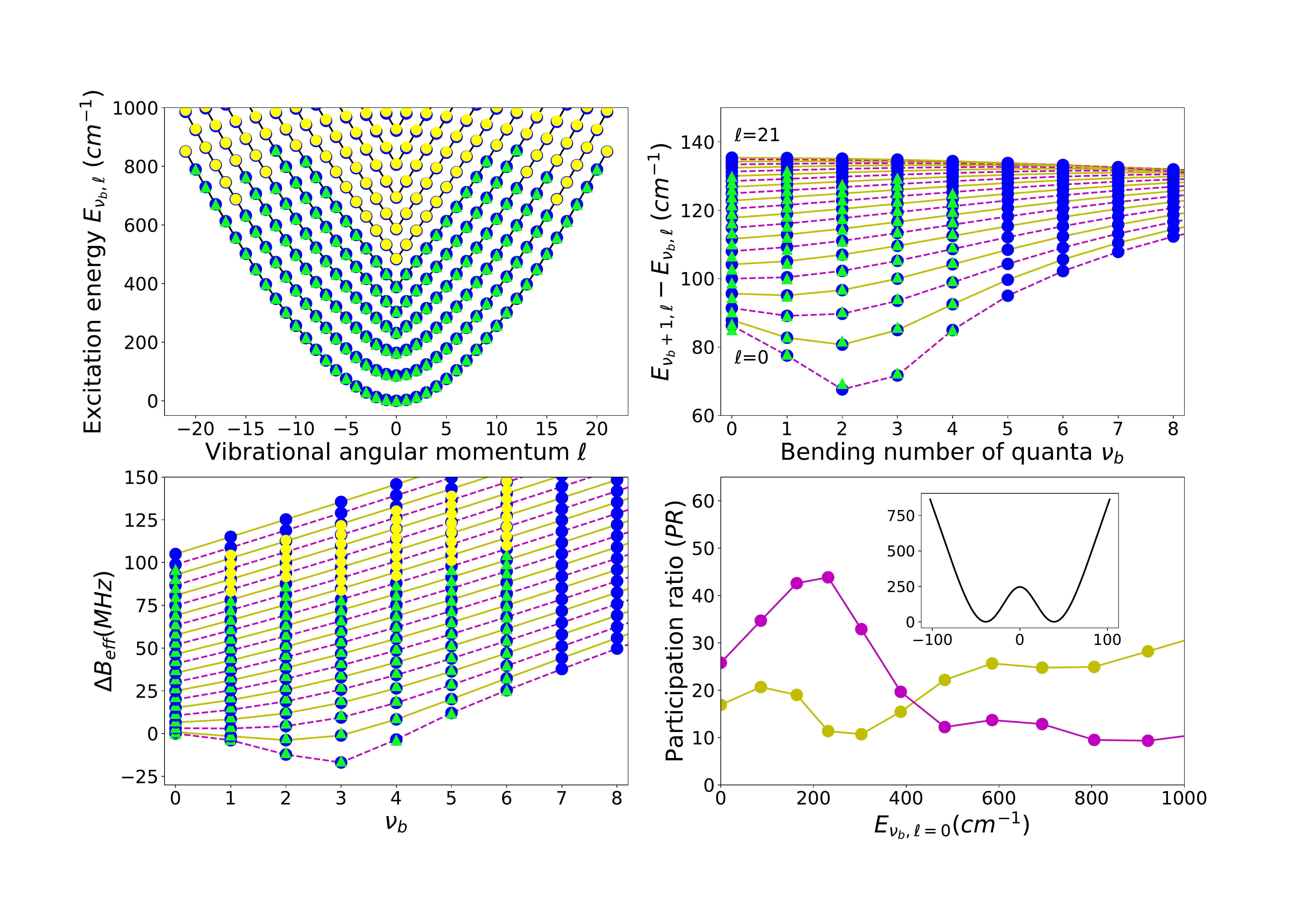}
            
            \caption{$\nu_7$ bending mode of NCNCS. Upper left panel: quantum
              monodromy plot. Upper right panel: Birge-Sponer plot (upper right
              panel) for the $\nu_7$ bending mode of NCNCS. Lower left panel:
              \(\Delta B_{eff}\) for as a function of the bending excitation
              \(\nu_b\). In these three panels blue circles (green triangles)
              are calculated (experimental) data. Yellow circles are predictions
              from \cite{Winnewisser2010}. Lower right panel: PR for $\ell = 0$
              eigenstates in the $U(2)$ (yellow line) and $SO(3)$ (magenta line)
              bases as a function of the state energy. The bending energy
              functional derived using the coherent state formalism is included
              as an inset in this panel.}
            \label{NCNCS}
        \end{figure}

    The reported and our calculated bending energies are included in
    Tab.~\ref{NCNCS_tb}. In Fig.~\ref{NCNCS} we show the quantum monodromy plot (upper
    left panel), the Birge-Sponer plot (upper right panel), the \(\Delta B_{eff}\)
    (lower left panel), and the PR (lower right panel) for the NCNCS \(\nu_7\)
    large amplitude bending mode.  In the \(\Delta B_{eff}\) case, a fit was
    performed making use of the expectation value of the number operator \(\hat
    n\) in the eigenfunctions resulting from the fit to the vibrational band
    origins and computing the values of the \(a_1\) and \(a_2\) parameters in
    (\ref{Beff}) that optimize the agreement with the observed \(\Delta B_{eff}\)
    values reported in Tables 8 to 11 of Ref.~\cite{Winnewisser2010}. The
    optimized parameter values are \(a_1 = 2.39(4)\)~MHz and \(a_2 =
    0.0108(4)\)~MHz with a fit having an \(rms = 1.29\)~MHz. We explored the
    effect of a cubic term in the expansion but it provides a marginal
    improvement and the two-parameter expansion (\ref{Beff}) already gives a fine
    result, achieving a satisfactory \(rms\).

    In Fig.~\ref{NCNCS}, the change from a quadratic to a linear pattern in the quantum monodromy
    plot and the location of the Dixon dip \cite{Dixon1964} in the Birge-Sponer plot
    indicate that the critical energy of the monodromy -or the ESQPT critical
    energy- is around the  $\nu_b=3$ overtone, as already discussed in the 
     literature \cite{Winnewisser2006,Winnewisser2005,Winnewisser2010,
       Larese2011}. In both plots the agreement between the reported data (green
     triangles) and
     the 2DVM results (blue circles) is good. We also include in the figure, as
     yellow circles, the values predicted by the GSRB model in
     Ref.~\cite{Winnewisser2010}. The agreement achieved for the \(\Delta
     B_{eff}\) is also very satisfactory.

    The PR (lower right panel) in Fig.~\ref{NCNCS}, makes evident the lack of
    significant localization effects in any basis for energies below
    the barrier. Nevertheless, the closest eigenstate to the critical
    energy --$\nu_b=3$ overtone-- is significantly more localized in
    the cylindrical oscillator basis, as predicted in recent works
    \cite{PB_Santos2016,Santos2015,Santos2016} for  states close to the
    critical energy of the ESQPT.
    
    The energy functional obtained making use of the coherent state
    approximation for this molecule, shown as an inset in the lower right panel
    of  Fig.~\ref{NCNCS}, provides an intuitive image for the potential associated to
    the model, with a low barrier to linearity and it allows for a rough
    estimation of the height of this barrier at $\sim 225$~cm$^{-1}$.
    Therefore,
    the different features shown in Fig.~\ref{NCNCS} confirm the nonrigid character of this molecule, which undergoes a bent-to-linear
    transition, in consonance with previous works \cite{Winnewisser2006,
      Winnewisser2005, Winnewisser2010, Larese2011}.

        \begin{table}[h]
        \centering
              \caption{Bending band origins and calculated energy term values and residuals for the  $\nu_7$ bending mode of NCNCS. Units of cm$^{-1}$.
              }
                \label{NCNCS_tb} 
            \begin{tabular}{cccc|cccc}

               \hline
                $(\nu_b,K)^a$ & $E_{origin}^b$ & Cal.$^c$ & $E_{origin}$-Cal.$^d$ & $(\nu_b,K)^a$ & $E_{origin}^b$ & Cal.$^c$ & $E_{origin}$-Cal.$^d$ \\
              \hline
( 0  1) &     3.42 &     3.3158 &   0.102   & ( 2  8) &   404.23 &   405.0085 &  -0.776   \\ 
( 0  2) &    13.38 &    13.0153 &   0.362   & ( 2  9) &   449.63 &   450.4394 &  -0.811   \\ 
( 0  3) &    29.24 &    28.5486 &   0.694   & ( 2 10) &   497.20 &   498.0187 &  -0.816   \\ 
( 0  4) &    50.32 &    49.2975 &   1.025   & ( 2 11) &   546.79 &   547.5617 &  -0.768   \\ 
( 0  5) &    75.99 &    74.6863 &   1.301   & ( 2 12) &   598.26 &   598.9047 &  -0.640   \\ 
( 0  6) &   105.70 &   104.2139 &   1.491   & ( 2 13) &   651.50 &   651.8997 &  -0.401   \\ 
( 0  7) &   139.03 &   137.4520 &   1.583   & ( 2 14) &   706.39 &   706.4114 &  -0.022   \\ 
( 0  8) &   175.61 &   174.0355 &   1.572   & ( 2 15) &   762.85 &   762.3142 &   0.532   \\ 
( 0  9) &   215.11 &   213.6505 &   1.464   & ( 3  0) &   232.26 &   231.4167 &   0.842   \\ 
( 0 10) &   257.30 &   256.0247 &   1.272   & ( 3  1) &   254.74 &   254.6472 &   0.089   \\ 
( 0 11) &   301.93 &   300.9194 &   1.015   & ( 3  2) &   283.01 &   283.2800 &  -0.267   \\ 
( 0 12) &   348.84 &   348.1224 &   0.716   & ( 3  3) &   315.44 &   315.9345 &  -0.493   \\ 
( 0 13) &   397.84 &   397.4436 &   0.399   & ( 3  4) &   351.30 &   351.9597 &  -0.658   \\ 
( 0 14) &   448.81 &   448.7108 &   0.096   & ( 3  5) &   390.15 &   390.9329 &  -0.785   \\ 
( 0 15) &   501.60 &   501.7664 &  -0.162   & ( 3  6) &   431.66 &   432.5398 &  -0.877   \\ 
( 0 16) &   556.12 &   556.4650 &  -0.343   & ( 3  7) &   475.60 &   476.5282 &  -0.926   \\ 
( 0 17) &   612.26 &   612.6711 &  -0.407   & ( 3  8) &   521.76 &   522.6870 &  -0.922   \\ 
( 0 18) &   669.94 &   670.2580 &  -0.318   & ( 3  9) &   569.99 &   570.8336 &  -0.845   \\ 
( 0 19) &   729.07 &   729.1059 &  -0.035   & ( 3 10) &   620.13 &   620.8062 &  -0.673   \\ 
( 0 20) &   789.59 &   789.1010 &   0.484   & ( 3 11) &   672.08 &   672.4592 &  -0.383   \\ 
( 1  0) &    85.04 &    86.2903 &  -1.253   & ( 3 12) &   725.71 &   725.6596 &   0.051   \\ 
( 1  1) &    90.10 &    91.2120 &  -1.112   & ( 4  0) &   304.64 &   303.0936 &   1.542   \\ 
( 1  2) &   103.60 &   104.4256 &  -0.830   & ( 4  1) &   340.46 &   339.6111 &   0.852   \\ 
( 1  3) &   123.67 &   124.1949 &  -0.528   & ( 4  2) &   377.14 &   376.8006 &   0.336   \\ 
( 1  4) &   149.02 &   149.2855 &  -0.270   & ( 4  3) &   415.88 &   415.9199 &  -0.042   \\ 
( 1  5) &   178.76 &   178.8451 &  -0.081   & ( 4  4) &   456.86 &   457.1720 &  -0.314   \\ 
( 1  6) &   212.29 &   212.2602 &   0.029   & ( 4  5) &   500.03 &   500.5241 &  -0.492   \\ 
( 1  7) &   249.13 &   249.0662 &   0.063   & ( 4  6) &   545.30 &   545.8758 &  -0.576   \\ 
( 1  8) &   288.93 &   288.8958 &   0.029   & ( 4  7) &   592.55 &   593.1078 &  -0.557   \\ 
( 1  9) &   331.39 &   331.4483 &  -0.056   & ( 4  8) &   641.68 &   642.0979 &  -0.421   \\ 
( 1 10) &   376.30 &   376.4701 &  -0.174   & ( 4  9) &   692.58 &   692.7267 &  -0.150   \\ 
( 1 11) &   423.44 &   423.7425 &  -0.301   & ( 4 10) &   745.16 &   744.8795 &   0.279   \\ 
( 1 12) &   472.66 &   473.0733 &  -0.412   & ( 4 11) &   799.34 &   798.4468 &   0.888   \\ 
( 1 13) &   523.81 &   524.2908 &  -0.479   & ( 4 12) &   855.03 &   853.3238 &   1.708   \\ 
( 1 14) &   576.77 &   577.2395 &  -0.469   & ( 5  0) &   389.60 &   388.0694 &   1.530   \\ 
( 1 15) &   631.43 &   631.7769 &  -0.351   & ( 5  1) &   433.35 &   432.1249 &   1.229   \\ 
( 2  0) &   162.94 &   163.8041 &  -0.864   & ( 5  2) &   476.53 &   475.7039 &   0.827   \\ 
( 2  1) &   173.03 &   173.9391 &  -0.910   & ( 5  3) &   520.61 &   520.1175 &   0.493   \\ 
( 2  2) &   192.71 &   193.5571 &  -0.849   & ( 5  4) &   566.10 &   565.8299 &   0.267   \\ 
( 2  3) &   218.52 &   219.2953 &  -0.773   & ( 5  5) &   613.16 &   612.9982 &   0.167   \\ 
( 2  4) &   248.97 &   249.6923 &  -0.717   & ( 5  6) &   661.86 &   661.6519 &   0.205   \\ 
( 2  5) &   283.24 &   283.9316 &  -0.695   & ( 5  7) &   712.16 &   711.7624 &   0.399   \\ 
( 2  6) &   320.77 &   321.4770 &  -0.703   & ( 5  8) &   764.04 &   763.2733 &   0.764   \\ 
( 2  7) &   361.20 &   361.9377 &  -0.735   & ( 5  9) &   817.44 &   816.1141 &   1.323   \\
       \hline
            \end{tabular}
            \flushleft
            {\footnotesize
              $^a$ Displaced oscillator basis quantum labels assigned to the
              optimized eigenvectors. \\
              $^b$ Bending band origins from \cite{Winnewisser2010}.\\
              $^c$ Calculated energies.\\
              $^d$ Difference between experimental and calculated energies.\\
            }
        \end{table}

\section{Summary and Conclusions}

We analyze the  bending vibrations of  four different molecules making use of the 2DVM most general Hamiltonian that includes up to four-body
interactions. The four molecular species  have been selected
trying to include examples of the different dynamics  associated with
the bending vibrational degree of freedom: linear, nonrigid, and bent. We present
optimized algebraic spectroscopic parameters for each one of the cases, as well
as the calculated bending spectrum and its comparison with reported values. The
reader can find in the Supplementary Material section the full spectrum, including not
yet measured or reported levels. We use the model energies wave functions to compute
the quantum monodromy, Birge-Sponer, and participation ratio for each case. In
the latter case we illustrate the eigenstate localization in the two basis
considered in the model.

Apart from its computational simplicity, one of the best 2DVM features is the
possibility of encompassing, within a simple model, the full gamut of bending
spectroscopic patterns that range from linear to bent, including the
feature-rich nonrigid cases. We have focused particularly in nonrigid cases,
where the bent-to-linear structural changes in the system, as it samples the top
of the barrier to linearity for increasing excitation energies, can be considered
as a perfect example of an ESQPT.

In a way, this work is a sequel of Refs.~\cite{Larese2011,Larese2013}, where a
systematic study of bending dynamics in molecular systems with and without ESQPT
signatures was performed for the first time. In our case, we use a higher-order
Hamiltonian to repeat the analysis, incorporating new reported data when
possible, trying to improve the results, and casting some light upon the modeled
physical systems.
The extension of the algebraic Hamiltonian to include three- and four-body
interactions has permitted us to model the available experimental data for the
four molecules considered and, according to the {\em rms} of the fits
(see Tab.~\ref{tab_results}), a satisfactory agreement is obtained between calculated
and reported energies.

In particular, in the HNC case, the inclusion in the fit of a single three-body
operator has dramatically improved the fit quality. As a bent molecule, we have
considered the bending spectrum of H$_2$S, where the coupling with the
rotational projection around the molecule-fixed $z$-axis in the Hamiltonian has
been included to grapple with the rotational contribution. The results obtained
in the fit to the H$_2$S bending levels largely improves previously published
results \cite{Larese2013}; with an accuracy such that our
predictions might be helpful for the assignment of new levels.

We have considered two nonrigid molecules, Si$_2$C and NCNCS. In the first case
we have obtained a fit within the experimental accuracy with only four
parameters using one- and two-body interactions. In the NCNCS case, the epitome
of a nonrigid molecule, we have also included results that reproduce rotational
spectrum patters associated with to the nonrigidity of the bending model. We
have computed \(\Delta B_{eff}\) obtaining in this case a satisfactory agreement
too.

We have included the mean field limit energy functional for the bending degree
of freedom of the nonrigid molecules under study using the intrinsic state
formalism. Given the level of abstraction of the algebraic model, developed far from the
traditional approach in phase space, this is a useful contribution as it
provides a more intuitive handle to the obtained results. One should always take
into consideration that this is a $1/N$ approximation, but still the bending
energy functional shed light on the potential shape, the height of the linearity
barriers, and the positions and number of minima.

Hence, we consider proved that the 4-body 2DVM Hamiltonian is a suitable
effective Hamiltonian for the analysis of bending vibrations and it provides new
venues to explore the ESQPT that occurs in the excitation spectrum of nonrigid
molecular species.  These results allow for an easier classification of the
bending degree of freedom among the possible situations existing between the
linear and bent limits, apart from being of great help in the assignment of
quantum labels to highly-excited bending states, often quite a cumbersome
task. Of course, one should always be aware of the model limits: it is a
phenomenological model needing a minimal set of values, either experimental or
extracted from experiment.

The programming codes used in this work are available upon request to the
authors, and they will be published soon. In the Supplementary Material section,
we provide predicted values for highly-excited bending levels of the molecules
studied in this work, with the expectation that they could be of help in the
measurement or assignment of experimental values. This is of particular
importance in the case of nonrigid molecular species, where an improved
knowledge of the critical energy region of the quantum monodromy --and,
therefore, of the ESQPT-- is of major importance and where we expect that our
approach could facilitate the assignment of quantum labels.

There are a number of developments that we are planning to undertake in a near
future, fostered by the success of the four-body Hamiltonian.  The full
description of the vibrational spectrum of a molecule implies the simultaneous
consideration of stretching and bending vibrational modes, as well as torsional,
rocking, or other large amplitude modes. This can be naturally accomplished in
the algebraic approach using coupled Lie algebras as shown in \cite{Larese2014}
for the case of two coupled benders or in \cite{SCastellanos2012,
  Lemus2014,Marisol2020,Ishikawa2002} in the case of coupling of a bending and two stretching
degrees of freedom. The latter works use an algebraic approach to obtain both
energy levels and spectrum line intensities, a very important step for a right
molecular characterization. The algebraic approach has performed very well in
the characterization of spectrum energies and line intensities of experimental
Franck-Condon \cite{Ishikawa2002, Muller:1998,Muller:1999,Muller:2000,
  Iachello:2000} and Raman intensities \cite{SCastellanos2012,
  Lemus2014,Marisol2020}. In this respect we are currently paying heed to the
modeling of highly excited bending progressions of HCN -with and without
stretching excitations- to facilitate the assignment in a spectrum complicated
by very large level density at high excitation energies.

In a different order, further developments of the model are currently being
considered based on the promising results for the transition state in the
isomerization of the [H,CN] system recently published in \cite{KRivera2019}. The
new developments include the possibility of extending the model to
simultaneously treat both isomer species. We will explore this model in
comparison with the results obtained using the GSRB model for this same system
\cite{Ross1983}; and looking for inspiration in other, more sophisticate models
as Refs.~\cite{Odaka2006, Mellau2008} and also
Refs.~\cite{Barnes2010,Barnes2011}, based on a formalism closer to the algebraic
one. In particular, we are planning to explore a configuration mixing formalism
akin to the one that has been successfully applied to nuclear systems
\cite{Duval1982}.

\paragraph{Acknowledgment}

We thank useful discussion with Profs.~Francesco Iachello, Renato Lemus, and
Georg Mellau. This project has received funding from the European Union's
Horizon 2020 research and innovation program under the Marie Sk\l odowska-Curie
grant agreement No 872081 and from the Spanish National Research, Development,
and Innovation plan (RDI plan) under the project PID2019-104002GB-C21.  This
work has also been partially supported by the Consejer\'{\i}a de Conocimiento,
Investigaci\'on y Universidad, Junta de Andaluc\'{\i}a and European Regional
Development Fund (ERDF), ref.  SOMM17/6105/UGR, by the Ministerio de Ciencia,
Innovaci\'on y Universidades (ref.COOPB20364), and by the Centro de Estudios
Avanzados en F\'{\i}sica, Matem\'aticas y Computaci\'on (CEAFMC) of the
University of Huelva.  JKR thanks support from the Youth Employment Initiative
and the Youth Guarantee program supported by the European Social Fund. FPB and
JKR thank the obtained support from the project UHU-1262561.  Computing
resources supporting this work were provided by the CEAFMC and Universidad de
Huelva High Performance Computer (HPC@UHU) located in the Campus Universitario
el Carmen and funded by FEDER/MINECO project UNHU-15CE-2848.


\appendix
\section{Operator Matrix elements}
\label{app-matrix_elements}

\subsubsection{Operator Matrix Elements in the Dynamical Symmetry (I)}
\label{sec-2-3-1}

The diagonal operators in this dynamical symmetry are

\begin{description}
\item[{Operator \(\hat n^p\)}]: \(\langle [N];n^\ell|\hat n^p|[N];n^\ell\rangle = n^p\) for \(p = 1,2,3,4\).

\item[{Operator \(\hat \ell^{2q}\)}]: \(\langle [N];n^\ell|\hat \ell^{2q}|[N];n^\ell\rangle = \ell^{2q}\) for \(q = 1,2\).

\item[{Operator \(\hat n^p\hat \ell^{2q}\)}]: \(\langle [N];n^\ell|\hat n^p\ell^{2q}|[N];n^\ell\rangle = n^p\ell^{2q}\) for \(p = 1,2\) and \(q = 1\).
\end{description}

The non-diagonal matrix elements in this basis are 

\begin{description}
\item[{\(SO(3)\) Casimir Operator \(\hat W^2\)}]: 
\end{description}

\begin{align}
\langle [N];n_2^\ell|{\hat W}^2|[N];n_1^\ell\rangle =&
\left[(N-n_1)(n_1+2)+(N-n_1+1)n_1 + \ell^2\right] \delta_{n_2,n_1} \nonumber\\
-& \sqrt{(N-n_1+2)(N-n_1+1)(n_1+\ell)(n_1-\ell)}\,\delta_{n_2,n_1-2} \nonumber\\
-&
\sqrt{(N-n_1)(N-n_1-1)(n_1+\ell+2)(n_1-\ell+2)}\,\delta_{n_2,n_1+2}~~.\nonumber
\end{align}

Note that this is the main non-diagonal operator in this case and it is
a band matrix as the non-zero matrix elements are located in the main and first diagonals only.

\begin{description}
\item[{Operator \(\hat n \hat W^2 + \hat W^2 \hat n\)}]: As the operator
\(\hat n\) is diagonal the matrix is also band diagonal with matrix
elements
\end{description}

{\footnotesize
\begin{align}
\langle [N];n_2^\ell|\hat n \hat W^2 + \hat W^2 \hat n|[N];n_1^\ell\rangle =&
2n_1\left[(N-n_1)(n_1+2)+(N-n_1+1)n_1 + \ell^2\right] \delta_{n_2,n_1} \nonumber\\
-& (2n_1-2)\sqrt{(N-n_1+2)(N-n_1+1)(n_1+\ell)(n_1-\ell)}\,\delta_{n_2,n_1-2}\nonumber\\
-& (2n_1+2)\sqrt{(N-n_1)(N-n_1-1)(n_1+\ell+2)(n_1-\ell+2)}\,\delta_{n_2,n_1+2}~~.\nonumber
\end{align}
}

\begin{description}
\item[{Operator \(\hat \ell^2 \hat W^2\)}]: This operator is computed for \(\ell\ne 0\) multiplying the \({\hat W}^2\) operator matrix times \(\ell^2\).

\item[{Operator \(\hat n^2 \hat W^2 + \hat W^2 \hat n^2\)}]: This is
computed as the \(\hat n \hat W^2 + \hat W^2 \hat n\) operator.
\end{description}

{\footnotesize
\begin{align}
\langle [N];n_2^\ell|\hat n^2 \hat W^2 + \hat W^2 \hat n^2|[N];n_1^\ell\rangle =&
2n_1^2\left[(N-n_1)(n_1+2)+(N-n_1+1)n_1 + \ell^2\right] \delta_{n_2,n_1} \nonumber\\
-& [n_1^2 + (n_1-2)^2]\sqrt{(N-n_1+2)(N-n_1+1)(n_1+\ell)(n_1-\ell)}\,\delta_{n_2,n_1-2}\nonumber\\
-& [n_1^2 + (n_1+2)^2]\sqrt{(N-n_1)(N-n_1-1)(n_1+\ell+2)(n_1-\ell+2)}\,\delta_{n_2,n_1+2}~~.\nonumber
\end{align}
}

\begin{description}
\item[{Operator \(\hat W^4\)}]: This operator is computed as the matrix
product of the \(\hat W^2\) operator matrix times itself.

\item[{Operator \(\hat W^2 \hat{\overline{W}}^2 + \hat{\overline{W}}^2\hat W^2 \)}]: In
this basis the only difference between the matrix elements of the
\(\hat W^2\) and \(\hat{\overline{W}}^2\) operators is the sign
of the non-diagonal contribution, which is positive in this
case. The full operator is computed via matrix multiplication.
\end{description}

\subsubsection{Operator Matrix Elements in the Dynamical Symmetry (II)}
\label{sec-2-3-2}

The diagonal operators in this dynamical symmetry are

\begin{description}
\item[{\(SO(3)\) Casimir Operator \(\hat W^2\)}]: \(\langle [N];\omega \ell|\hat W^2|[N]; \omega \ell\rangle = \omega(\omega + 1)\).

\item[{Operator \(\hat \ell^{2q}\)}]: \(\langle [N];\omega \ell|\hat \ell^{2q}|[N];\omega \ell\rangle = \ell^{2q}\) for \(q = 1,2\).

\item[{Operator \(\hat \ell^2 \hat W^2\)}]: \(\langle [N];\omega \ell|\hat \ell^2 \hat W^2|[N]; \omega \ell\rangle = \ell^2 \omega(\omega + 1)\).

\item[{Operator \(\hat W^4\)}]: \(\langle [N];\omega \ell|\hat W^4|[N]; \omega \ell\rangle = \omega^2(\omega + 1)^2\).
\end{description}

The non-diagonal matrix elements in this basis are 
\begin{description}
\item[{Operator \(\hat n\)}]: 
\end{description}

{\footnotesize
\begin{align}
  \langle [N];w_2^\ell|\hat n|[N];w_1^\ell\rangle =&
  \left\{\frac{(N-w_1)\left[(w_1-\ell+2)(w_1-\ell+1) +
        (w_1+\ell+2)(w_1+\ell+1)\right]}{2(2w_1+1)(2w_1+3)} \right. \nonumber\\
  & + \left.\frac{(N+w_1+1)\left[(w_1+\ell)(w_1+\ell-1) +
        (w_1-\ell)(w_1-\ell-1)\right]}{2(2w_1+1)(2w_1-1)} \right\}\,
  \delta_{w_2,w_1} \nonumber\\
+& \sqrt{\frac{(N-w_1)(N+w_1+3)(w_1-\ell+2)(w_1-\ell+1)(w_1+\ell+2)(w_1+\ell+1)}{(2w_1+1)(2w_1+3)^2(2w_1+5)}}\delta_{w_2,w_1+2}\nonumber\\
+& \sqrt{\frac{(N-w_1+2)(N+w_1+1)(w_1-\ell)(w_1-\ell-1)(w_1+\ell)(w_1+\ell-1)}{(2w_1-3)(2w_1-1)^2(2w_1+1)}}\delta_{w_2,w_1-2}\nonumber
\end{align}
}

Note that this is the main non-diagonal operator in this case and it is again
a band matrix with non-zero matrix elements located in the main and first diagonals only. The \(\hat n\) matrix element in this basis are taken from
\cite{PBernal2008} with a typo that has been corrected.
\begin{description}
\item[{Operators \(\hat n^2\), \(\hat n^3\), and \(\hat n^4\)}]: These three operators are computed by matrix multiplication of the basic operator ${\hat n}$.

\item[{Operator \(\hat n \hat \ell^{2}\)}]: This operator is computed for \(\ell\ne 0\) multiplying the \({\hat n}\) operator matrix times \(\ell^2\).

\item[{Operator \(\hat n \hat W^2 + \hat W^2 \hat n\)}]: As the operator
\(\hat n\) is band diagonal \(\hat W^2\) is diagonal this operator matrix is also band diagonal with matrix
elements
\end{description}

{\footnotesize
\begin{align}
\langle [N];w_2 \ell|\hat n \hat W^2 + \hat W^2 \hat n|[N];w_1 \ell\rangle &=
 2\omega_1(\omega_1+1)  \left\{\frac{(N-w_1)\left[(w_1-\ell+2)(w_1-\ell+1) +
        (w_1+\ell+2)(w_1+\ell+1)\right]}{2(2w_1+1)(2w_1+3)} \right. \nonumber\\
  & + \left.\frac{(N+w_1+1)\left[(w_1+\ell)(w_1+\ell-1) +
        (w_1-\ell)(w_1-\ell-1)\right]}{2(2w_1+1)(2w_1-1)} \right\}\,
  \delta_{w_2,w_1} \nonumber\\
+& \left[\omega_1(\omega_1+1) + (\omega_1 +2)(\omega_1+3)\right]\nonumber\\
\times &  \sqrt{\frac{(N-w_1)(N+w_1+3)(w_1-\ell+2)(w_1-\ell+1)(w_1+\ell+2)(w_1+\ell+1)}{(2w_1+1)(2w_1+3)^2(2w_1+5)}}\delta_{w_2,w_1+2}\nonumber\\
+& \left[(\omega_1 - 2)(\omega_1-1) + \omega_1 (\omega_1+1)\right]\nonumber\\
\times & \sqrt{\frac{(N-w_1+2)(N+w_1+1)(w_1-\ell)(w_1-\ell-1)(w_1+\ell)(w_1+\ell-1)}{(2w_1-3)(2w_1-1)^2(2w_1+1)}}\delta_{w_2,w_1-2}\nonumber
\end{align}
}

\begin{description}
\item[{Operator \(\hat n^2 \hat W^2 + \hat W^2 \hat n^2\)}]: This is
computed in the same way that the \(\hat n \hat W^2 + \hat W^2
     \hat n\) operator but taking into account that the \(\hat n^2\)
operator is double banded. Therefore the operator matrix elements
can be expressed as follow
\end{description}

\begin{align}
\langle [N];w_2 \ell|\hat n^2 \hat W^2 + \hat W^2 \hat n^2|[N];w_1 \ell\rangle &=
 2\omega_1(\omega_1+1) [\hat n^2]_{w_1,w_1} ~ \delta_{w_2,w_1} \nonumber\\
+& \left[\omega_1(\omega_1+1) + (\omega_1 +2)(\omega_1+3)\right][\hat n^2]_{w_1,w_1+2} ~ \delta_{w_2,w_1+2}\nonumber\\
+& \left[(\omega_1 - 2)(\omega_1-1) + \omega_1 (\omega_1+1)\right][\hat n^2]_{w_1,w_1-2} ~ \delta_{w_2,w_1-2}\nonumber\\
+& \left[\omega_1(\omega_1+1) + (\omega_1 +4)(\omega_1+5)\right][\hat n^2]_{w_1,w_1+4} ~ \delta_{w_2,w_1+4}\nonumber\\
+& \left[(\omega_1 - 4)(\omega_1-3) + \omega_1 (\omega_1+1)\right][\hat n^2]_{w_1,w_1-4} ~ \delta_{w_2,w_1-4}\nonumber~,
\end{align}
\noindent where \([\hat n^2]_{w_i,w_j}\) are the \(\hat n^2\) operator matrix elements.

\begin{description}
\item[{Operator \(\hat W^2 \hat{\overline{W}}^2 + \hat{\overline{W}}^2\hat W^2 \)}]: In
this basis we need first to compute the matrix elements of the \(\hat{\overline{W}}^2\) making use of Eqs. (37) and (38) of Ref. \cite{PBernal2008}.
\end{description}
\begin{equation}\nonumber
\langle [N];w_2 \ell_2|\hat R_{-}|[N];w_1 \ell_1\rangle = A_{w_1,\ell_1} \, \delta_{w_2,w_1} ~ \delta_{\ell_2,\ell_1-1} + B_{w_1,\ell_1} \, \delta_{w_2,w_1-2} ~ \delta_{\ell_2,\ell_1-1} + C_{w_1,\ell_1} \, \delta_{w_2,w_1+2} ~ \delta_{\ell_2,\ell_1-1}~,
\end{equation}
\noindent where 
\begin{align}
A_{w,\ell} = & \frac{(2N+3)(2\ell + 1)}{(2w-1)(2w+3)} \sqrt{(w+\ell)(w-\ell+1)/2}\nonumber\\ 
B_{w,\ell} = & -\sqrt{\frac{2(N+w+1)(N-w+2)(w+\ell)(w-\ell)(w+\ell-1)(w+\ell-2)}{(2w+1)(2w-1)^2(2w-3)}}\nonumber\\ 
C_{w,\ell} = & \sqrt{\frac{2(N+w+3)(N-w)(w+\ell+1)(w-\ell+1)(w-\ell+2)(w-\ell+3)}{(2w+1)(2w+3)^2(2w+5)}}\nonumber~.
\end{align}

The previous result can be used for the derivation of an expression for the \(\hat R_{+}\) operator matrix elements
\begin{align}
\langle [N];w_2 \ell_2|\hat R_{-}|[N];w_1 \ell_1\rangle^\dagger =& \langle [N];w_1 \ell_1|\hat R_{+}|[N];w_2 \ell_2\rangle \nonumber \\
=& A_{w_2,\ell_2+1} \, \delta_{w_1,w_2} ~ \delta_{\ell_1,\ell_2+1} + B_{w_2+2,\ell_2+1} \, \delta_{w_1,w_2-2} ~ \delta_{\ell_1,\ell_2+1}  \nonumber \\ & + C_{w_2-2,\ell_2+1} \, \delta_{w_1,w_2+2} ~ \delta_{\ell_1,\ell_2+1}~.\nonumber
\end{align}

The upper diagonal matrix elements of the Casimir operator \(\hat{\overline{W}}^2 = \hat R_{+}\hat R_{-} + \hat \ell^2\) can then be expressed as
\begin{align}
\langle [N];w_2 \ell|\hat{\overline{W}}^2|[N];w_1 \ell\rangle &=
( A^2_{w_1,\ell_1} + B^2_{w_1,\ell_1} + C^2_{w_1,\ell_1} ) \, \delta_{w_2,w_1} \nonumber\\
+& (A_{w_1,\ell} \, B_{w_1+2,\ell} + C_{w_1,\ell} \, A_{w_1+2,\ell})\, \delta_{w_2,w_1+2}\nonumber\\
+& C_{w_1,\ell} \, B_{w_1+4,\ell} \, \delta_{w_2,w_1+4}\nonumber~,
\end{align}
\noindent and the lower diagonal matrix elements can be computed considering that the upper and lower bandwidths are the same.

The \(\hat W^2 \hat{\overline{W}}^2 + \hat{\overline{W}}^2\hat W^2 \) operator is then computed as for the $\hat n^2 \hat W^2 + \hat W^2 \hat n^2$ operator 

\begin{align}
\langle [N];w_2 \ell|\hat W^2 \hat{\overline{W}}^2 + \hat{\overline{W}}^2\hat W^2|[N];w_1 \ell\rangle &=
 2\omega_1(\omega_1+1) \, [\hat{\overline{W}}^2]_{w_1,w_1} ~ \delta_{w_2,w_1} \nonumber\\
+& \left[\omega_1(\omega_1+1) + (\omega_1 +2)(\omega_1+3)\right] \, [\hat {\overline{W}}^2]_{w_1,w_1+2} ~ \delta_{w_2,w_1+2}\nonumber\\
+& \left[(\omega_1 - 2)(\omega_1-1) + \omega_1 (\omega_1+1)\right] \, [\hat {\overline{W}}^2]_{w_1,w_1-2} ~ \delta_{w_2,w_1-2}\nonumber\\
+& \left[\omega_1(\omega_1+1) + (\omega_1 +4)(\omega_1+5)\right] \, [\hat {\overline{W}}^2]_{w_1,w_1+4} ~ \delta_{w_2,w_1+4}\nonumber\\
+& \left[(\omega_1 - 4)(\omega_1-3) + \omega_1 (\omega_1+1)\right] \, [\hat {\overline{W}}^2]_{w_1,w_1-4} ~ \delta_{w_2,w_1-4}\nonumber~,
\end{align}
\noindent where \([\hat{\overline W}^2]_{w_i,w_j}\) are the \(\hat{\overline W}^2\) operator matrix elements. 
\section{Coherent state approach results}
\label{app_coherent}
The matrix elements of the different operators in Hamiltonian  (\ref{H4b}),
\(\expval{\hat O}_{c.s.} = \bra{[N];\textbf{r}}\hat O\ket{[N];\textbf{r}}\) for
one- to four-body operators are
\begin{itemize}
\item \textbf{One-body operator:}
  \begin{itemize}
  \item[o] $\expval{\hat n}_{c.s.} = N \frac{r^2}{1+r^2}$
  \end{itemize}
\item \textbf{Two-body operators:}
  \begin{itemize}
  \item[o]
    $\expval{\hat{n}^2}_{c.s.} =N \frac{r^2}{1+r^2} +N(N-1)
    \frac{r^4}{\left(1+r^2\right)^2}$
  \item[o] $\expval{\hat{\ell}^2}_{c.s.} = \expval{\hat n}_{c.s.}$
  \item[o]
    $\expval{\hat{W}^2}_{c.s.} = 2N + N(N-1)\frac{4r^2}{(1+r^2)^2}$
  \end{itemize}
\item \textbf{Three-body operators:}
  \begin{itemize}
  \item[o]
    $\expval{\hat{n}^3}_{c.s.} = N \frac{r^2}{1+r^2} +3N(N-1)
    \frac{r^4}{\left(1+r^2\right)^2} +
    N(N-1)(N-2)\frac{r^6}{(1+r^2)^3}$
  \item[o]
    $\expval{\hat{n}\hat{\ell}^2}_{c.s.} = \expval{\hat{n}^2}_{c.s.} $
  \item[o]
    \raisebox{-0.7\baselineskip}{$\begin{array}{ll}\expval{\hat{n}\hat{W}^2+\hat{W}^2\hat{n}}_{c.s.}
        & = 4N\frac{r^2}{1+r^2} + 4N(N-1)\frac{r^4}{(1+r^2)^2} \\ & +
        12N(N-1)\frac{r^2}{(1+r^2)^2}+8N(N-1)(N-2)\frac{r^4}{(1+r^3)^3}\end{array}$}
  \end{itemize}
\item \textbf{Four-body operators:}
  \begin{itemize}
  \item[o]
    \raisebox{-0.7\baselineskip}{
      \(
      \begin{array}{ll}
        \expval{\hat{n}^4}_{c.s.} & = N\frac{r^2}{1+r^2} + 7N(N-1)\frac{r^4}{(1+r^2)^2} + 6N(N-1)(N-2)\frac{r^6}{(1+r^2)^3} \\ 
                                  & + N(N-1)(N-2)(N-3)\frac{r^8}{{(1+r^2)}^4}
      \end{array}\)
    }
  \item[o]
    $\expval{\hat{n}^2\hat{\ell}^2}_{c.s.} = \expval{\hat{n}^3}_{c.s.}$
  \item[o]
    $\expval{\hat{\ell}^4}_{c.s.} = N\frac{r^2}{1+r^2} +
    3N(N-1)\frac{r^4}{(1+r^2)^2}$
  \item[o]
    $\expval{\hat{\ell}^2\hat{W}^2}_{c.s.} =
    2N\frac{r^2}{1+r^2}+4N(N-1)\frac{r^4+r^2}{(1+r^2)^2}+4N(N-1)(N-2)\frac{r^4}{(1+r^2)^3}$
  \item[o]
    \raisebox{-0.7\baselineskip}{$\begin{array}{ll}\expval{\hat{n}^2\hat{W}^2+\hat{W}^2\hat{n}^2}_{c.s.}
        & = 4N\frac{r^2}{1+r^2}+N(N-1)\frac{12r^4+16r^2}{(1+r^2)^2} \\
        & + N(N-1)(N-2)\frac{4r^6+28r^4}{(1+r^2)^3} +
        N(N-1)(N-2)(N-3)\frac{8r^6}{(1+r^2)^4}\end{array}$}
  \item[o]
    \raisebox{-0.7\baselineskip}{$\begin{array}{ll}\expval{\hat{W}^4}_{c.s.}
        & = 4N(2N-1) +
        24N(N-1)\frac{r^2}{(1+r^2)^2}+32N(N-1)(N-2)\frac{r^4+r^2}{(1+r^2)^3}
        \\ & +16N(N-1)(N-2)(N-3)\frac{r^4}{(1+r^2)^4} \end{array}$}
  \item[o]
    \raisebox{-0.7\baselineskip}{$\begin{array}{ll}\frac{1}{2}\expval{\hat{W}^2\hat{\overline{W}}^2+\hat{\overline{W}}^2\hat{W}^2}_{c.s.}
        & =
        4N+N(N-1)\frac{4r^4+28r^2}{(1+r^2)^2}\\&+8N(N-1)(N-2)\frac{r^4+r^2}{(1+r^2)^3}\end{array}$}
  \end{itemize}
  
  The connection of the energy functional with results obtained using the
  traditional analysis in phase space is hindered by the need of a connection
  between the unitless $r$ variable and the physical bending coordinate. A
  connection has been worked out for the two dynamical symmetries
  \cite{PBernal2005, Larese2011}, with a simple linear relationship based on the
  molecular G matrix elements. However, it is necessary to look for a more
  elaborate relationship between the classical coordinate resulting from the
  coherent state approach and the physical coordinate valid for more general
  cases. Still, we can obtain a qualitative description of the system potential
  by directly connecting the $r$ intrinsic approach classical variable to the
  deviation of linearity angle,
  $ \theta(rad) = \frac{\theta_e (rad)} {r_{min}}r$, making use of the
  experimental information on the molecular equilibrium structure. We have used
  the intrinsic state approach only in the two nonrigid molecules, Si$_2$C and
  NCNCS. In this way the energy functional has the minimum located at the right
  position, and the figure offers a pictorial and intuitive perspective to the
  obtained results and a qualitative image of the potential shape. We are
  planning to work in a more involved scaling in a future work.
  
\end{itemize}




\newpage
\bibliography{mybibfile.bib}
\bibliographystyle{elsarticle-num}

\end{document}